\documentclass[longauth]{aa} 
\usepackage{graphicx}
\usepackage{txfonts}
\usepackage{natbib}
\bibpunct{(}{)}{;}{a}{}{,}
\def\s-1{\rm {s^{-1}}}

\usepackage{url}

\begin{document}

   \title{Studying Galactic interstellar turbulence through fluctuations in synchrotron emission}

   \subtitle{First LOFAR Galactic foreground detection.}

   \titlerunning{The first diffuse Galactic foreground detection with LOFAR}

   \author{M.~Iacobelli\inst{1}\inst{2}
          \and
          M.~Haverkorn\inst{3,1}
          \and
          E.~Orr\'u\inst{2,3}
	  \and
          R.~F.~Pizzo\inst{2}
          \and J.~Anderson\inst{4}
          \and R.~Beck\inst{4}
          \and M.~R.~Bell\inst{5}
          \and A.~Bonafede\inst{6}
          \and K.~Chyzy\inst{7}
          \and R.-J.~Dettmar\inst{8}
          \and T.A.~En{\ss}lin\inst{9}
          \and G.~Heald\inst{2}
          \and C.~Horellou\inst{10}
          \and A.~Horneffer\inst{4}
          \and W.~Jurusik\inst{7}
          \and H.~Junklewitz\inst{9}
          \and M.~Kuniyoshi\inst{4}
          \and D.~D.~Mulcahy\inst{4}
          \and R.~Paladino\inst{35}
          \and W.~Reich\inst{4}
          \and A.~Scaife\inst{11}
          \and C.~Sobey\inst{4}
          \and C.~Sotomayor-Beltran\inst{12}
          \and A.~Alexov\inst{13}
          \and A.~Asgekar\inst{2} 
          \and I.~M.~Avruch\inst{14}
          \and M.~E.~Bell\inst{15}
          \and I.~van Bemmel\inst{2}
          \and M.~J.~Bentum\inst{2}
          \and G.~Bernardi\inst{16}
          \and P.~Best\inst{17}
          \and L.~B{\i}rzan\inst{1}
          \and F.~Breitling\inst{18}
          \and J.~Broderick\inst{11}
          \and W.~N.~Brouw\inst{19}
          \and M.~Br\"uggen\inst{6}
          \and H.~R.~Butcher\inst{20}
          \and B.~Ciardi\inst{5}
          \and J.~E.~Conway\inst{10}
          \and F.~de Gasperin\inst{6}
          \and E.~de Geus\inst{2}
          \and S.~Duscha\inst{2}
          \and J.~Eisl\"offel\inst{21}
          \and D.~Engels\inst{22}
          \and H.~Falcke\inst{3}\inst{2}
          \and R.~A.~Fallows\inst{2}
          \and C.~Ferrari\inst{23}
          \and W.~Frieswijk\inst{2}
          \and M.~A.~Garrett\inst{2}\inst{1}
          \and J.~Grie{\ss}meier\inst{24}
          \and A.~W.~Gunst\inst{2}
          \and J.~P.~Hamaker\inst{2}
          \and T.~E.~Hassall\inst{11}\inst{29}
          \and J.~W.~T.~Hessels\inst{2,31}
          \and M.~Hoeft\inst{21}
          \and J.~H\"orandel\inst{3}
          \and V.~Jelic\inst{2}
          \and A.~Karastergiou\inst{25}
          \and V.~I.~Kondratiev\inst{2,32}
          \and L.~V.~E.~Koopmans\inst{19}
          \and M.~Kramer\inst{4}
          \and G.~Kuper\inst{2}
          \and J.~van Leeuwen\inst{2}
          \and G.~Macario\inst{23}
          \and G.~Mann\inst{18}
          \and J.~P.~McKean\inst{2}
          \and H.~Munk\inst{2}
          \and M.~Pandey-Pommier\inst{26}
          \and A.~G.~Polatidis\inst{2}
          \and H.~R\"ottgering\inst{1}
          \and D.~Schwarz\inst{27}
          \and J.~Sluman\inst{2}
          \and O.~Smirnov\inst{28,33}
          \and B.~W.~Stappers\inst{29}
          \and M.~Steinmetz\inst{18}
          \and M.~Tagger\inst{24}
          \and Y.~Tang\inst{2}
          \and C.~Tasse\inst{30}
          \and C.~Toribio\inst{2}
          \and R.~Vermeulen\inst{2}
          \and C.~Vocks\inst{18}
          \and C.~Vogt\inst{2}
          \and R.~J.~van Weeren\inst{16}
          \and M.~W.~Wise\inst{2,31}
          \and O.~Wucknitz\inst{34,4}
          \and S.~Yatawatta\inst{2}
          \and P.~Zarka\inst{30}
          \and A.~Zensus\inst{4}
   }
   \offprints{M. Iacobelli}

   \institute{
   Leiden Observatory, Leiden University, PO Box 9513, 2300 RA Leiden, The Netherlands\\
              \email{iacobelli@strw.leidenuniv.nl} 
  \and Netherlands Institute for Radio Astronomy (ASTRON), Postbus 2, 7990 AA Dwingeloo, The Netherlands
  \and Radboud University Nijmegen, Heijendaalseweg 135, 6525 AJ Nijmegen, the Netherlands
  \and Max-Planck-Institut f\"ur Radioastronomie, Auf dem H\"ugel 69, 53121 Bonn, Germany
  \and Max Planck Institute for Astrophysics, Karl Schwarzschild Str. 1, 85741 Garching, Germany
  \and University of Hamburg, Gojenbergsweg 112, 21029 Hamburg, Germany
  \and Jagiellonian University, ul. Orla 171, PL-30244, Krak\'ow, Poland
  \and Astronomisches Institut der Ruhr-Universit\"at Bochum, Universitaetsstrasse 150, 44780 Bochum, Germany
  \and Max-Planck Institute for Astrophysics, Karl-Schwarzschild-Strasse 1, 85748 Garching bei M\"unchen, Germany
  \and Onsala Space Observatory, Dept. of Earth and Space Sciences, Chalmers University of Technology, SE-43992 Onsala, Sweden
  \and School of Physics and Astronomy, University of Southampton, Southampton, SO17 1BJ, UK
  \and Astronomisches Institut der Ruhr-Universit\"at Bochum, Universitaetsstrasse 150, 44780 Bochum, Germany
  \and Space Telescope Science Institute, 3700 San Martin Drive, Baltimore, MD 21218, USA
  \and SRON Netherlands Insitute for Space Research, PO Box 800, 9700 AV Groningen, The Netherlands
  \and ARC Centre of Excellence for All-sky astrophysics (CAASTRO), Sydney Institute of Astronomy, University of Sydney Australia
  \and Harvard-Smithsonian centre for Astrophysics, 60 Garden Street, Cambridge, MA 02138, USA
  \and Institute for Astronomy, University of Edinburgh, Royal Observatory of Edinburgh, Blackford Hill, Edinburgh EH9 3HJ, UK
  \and Leibniz-Institut f\"ur Astrophysik Potsdam (AIP), An der Sternwarte 16, 14482 Potsdam, Germany
  \and University of Groningen, Kapteyn Astronomical Institute, PO Box 800, 9700 AV Groningen, The Netherlands
  \and Research School of Astronomy and Astrophysics, Australian National University, Mt Stromlo Obs., via Cotter Road, Weston, A.C.T. 2611, Australia
  \and Th\"uringer Landessternwarte, Sternwarte 5, D-07778 Tautenburg, Germany
  \and Hamburger Sternwarte, Gojenbergsweg 112, D-21029 Hamburg
  \and Laboratoire Lagrange, UMR7293, Universit\`e de Nice Sophia-Antipolis, CNRS, Observatoire de la C\'{o}te d'Azur, 06300 Nice, France
  \and Laboratoire de Physique et Chimie de l' Environnement et de l' Espace, LPC2E UMR 7328 CNRS, 45071 Orl\'{e}ans Cedex 02, France
  \and Astrophysics, University of Oxford, Denys Wilkinson Building, Keble Road, Oxford OX1 3RH
  \and Centre de Recherche Astrophysique de Lyon, Observatoire de Lyon, 9 av Charles Andr\'{e}, 69561 Saint Genis Laval Cedex, France
  \and Fakult\"at f\"ur Physik, Universit\"{a}t Bielefeld, Postfach 100131, D-33501, Bielefeld, Germany
  \and Centre for Radio Astronomy Techniques \& Technologies (RATT), Department of Physics and Electronics, Rhodes University, PO Box 94, Grahamstown 6140, South Africa
  \and Jodrell Bank centre for Astrophysics, School of Physics and Astronomy, The University of Manchester, Manchester M13 9PL,UK
  \and LESIA, UMR CNRS 8109, Observatoire de Paris, 92195 Meudon, France
  \and Astronomical Institute 'Anton Pannekoek', University of Amsterdam, Postbus 94249, 1090 GE Amsterdam, The Netherlands
  \and Astro Space centre of the Lebedev Physical Institute, Profsoyuznaya str. 84/32, Moscow 117997, Russia
  \and SKA South Africa, 3rd Floor, The Park, Park Road, Pinelands, 7405, South Africa
  \and Argelander-Institut f\"{u}r Astronomie, University of Bonn, Auf dem H\"{u}gel 71, 53121, Bonn, Germany
  \and Universit\'{a} di Bologna - INAF ALMA regional centre, Via P. Gobetti 101, I-40129, Bologna, Italy 
  }
  
   \date{Received: 04 June 2013 / Accepted: 17 July 2013 }
 
  \abstract
   {}
{The characteristic outer scale of turbulence (i.e.\ the scale at which the dominant source of turbulence injects energy to the  interstellar medium) and the ratio of the random to ordered components of the magnetic field are key parameters to characterise magnetic turbulence in the interstellar gas, which affects the propagation of cosmic rays within the Galaxy. We provide new constraints to those two parameters.}
{We use the LOw Frequency ARray (LOFAR) to image the diffuse continuum emission in the Fan region at $(l,b)\sim(137.0^{\circ},+7.0^{\circ})$ at $80''\times70''$ resolution in the range [146,174]~MHz. We detect multi-scale fluctuations in the Galactic synchrotron emission and compute their power spectrum. Applying theoretical estimates and derivations from the literature for the first time, we derive the outer scale of turbulence and the ratio of random to ordered magnetic field from the characteristics of these fluctuations .}
{We obtain the deepest image of the Fan region to date and find diffuse continuum emission within the primary beam. The power spectrum displays a power law behaviour for scales between 100 and 8~arcmin with a  slope $\alpha=-1.84\pm 0.19$. We find an upper limit of $\sim 20$~pc for the outer scale of the magnetic interstellar turbulence toward the Fan region, which is in agreement with previous estimates in literature. We also find a variation of the ratio of random to ordered field as a function of Galactic coordinates, supporting different turbulent regimes.}
{We present the first LOFAR detection and imaging of the Galactic diffuse synchrotron emission around 160~MHz from the highly polarized Fan region. The power spectrum of the foreground synchrotron fluctuations is approximately a power law with a slope $\alpha\approx-1.84$ up to angular multipoles of $\lesssim1300$, corresponding to an angular scale of $\sim8$~arcmin. We use power spectra fluctuations from LOFAR as well as earlier GMRT and WSRT observations to constrain the outer scale of turbulence ($L_{out}$) of the Galactic synchrotron foreground, finding a range of plausible values of $10-20$~pc. Then, we use this information to deduce lower limits of the ratio of ordered to random magnetic field strength. These are found to be 0.3, 0.3, and 0.5 for the LOFAR, WSRT and GMRT fields considered respectively. Both these constraints are in agreement with previous estimates.}

\maketitle

\section{Introduction}
\label{s:intro}

The Galactic interstellar medium (ISM) is a complex and diffuse thermodynamic system with physical properties such as temperature and density spanning many orders, which define three main phases: the ``hot'', the ``warm'', and the ``cold'' phase. Moreover, the ISM is both magnetised and turbulent. Many efforts have been made over the past decades to characterise the magnetic fields and the turbulence in the ISM as well as their mutual dependence. However, fundamental parameters regarding both the Galactic magnetic field structure (e.g.\ the number and spatial location of large-scale reversals, the structure in the halo) and turbulence (e.g.\ the physical scale of energy injection, the sonic and Alfv\'enic Mach numbers) are still poorly constrained.

In this paper, we focus on the interplay of the Galactic magnetic field with turbulence in the ISM by estimating the physical scale of energy injection, $L_{out}$. This parameter defines the largest linear scale of the turbulent component of the Galactic magnetic field. Towards high Galactic latitudes an injection scale of about $140$~pc is found by \citet{Chepurnov10}, who were studying the velocity spectrum of the 21~cm line. Using structure functions of rotation measures, \citet{OhnoShibata93} found a large $L_{out}\lesssim100$~pc when averaging over large parts of the sky. \citet{Haverkorn08} confirmed this large outer scale for interarm regions in the Galactic plane using the same method; however, they found a much smaller outer scale $L_{out}\lesssim10$~pc in the spiral arms. This is in agreement with \citet{Clegg92}, who quote values of $0.1-10$~pc in the Galactic disk mostly towards the Sagittarius arm. Also, arrival anisotropies in TeV cosmic ray (CR) nuclei can be best explained by a magnetised, turbulent ISM on a maximum scale of about 1~pc \citep{Malkov10}.

In principle, one could expect multiple scales of energy injection in the ISM \citep{NotaKatgert10}. However, \citet[][]{MacLow04} showed from energy arguments that supernova remnants are expected to be the dominant energy source of the turbulence. Instead, the wide range of estimates of $L_{out}$ can be explained by a non-uniform spatial distribution of sources powering turbulence at the same scale of energy injection \citep[see e.g.][]{Haverkorn08}.  
In addition, the typical linear scale of turbulent regions in the ISM is an important parameter in the modelling of CR propagation. Anisotropies in the distribution of Galactic CR arrival directions on the sky have been measured by several experiments both on large (i.e.\ dipolar anisotropy) and small (i.e. between $10^{\circ}-30^{\circ}$) scales in the TeV-PeV energy range. Anisotropic magneto-hydrodynamic (MHD) turbulence in the interstellar magnetic field has also been proposed to explain such large-scale \citep{Battaner09} and small-scale \citep{Malkov10} anisotropies in the CR arrival directions at Earth. Recently \citet{Giacinti12a} have proposed the observed anisotropies to be the result of the scattering of TeV-PeV CR across the local magnetic turbulence, and thus within a few tens of parsecs from Earth.

Different observational methods and tracers can be used to study the properties of turbulence and/or magnetic fields in the ISM \citep[see e.g.][]{Elmegreen04,Scalo04} because they affect both the particle density as well as the emission, absorption, and propagation of radiation. Most of the observations about large-scale Galactic magnetic fields rely on Faraday rotation measures (RMs), where the imprints of magnetic fields and thermal electron density fluctuations are mixed; therefore, RM data allow direct study of fluctuations in the Galactic magnetic field only with a reliable electron density model. But the  radio synchrotron continuum of our Galaxy should also contain imprints of the magnetised turbulence in the ISM \citep[see e.g.][]{Eilek89a,Eilek89b,Waelkens09,Junklewitz11,Lazarian12}. Below $\nu \lesssim 1$~GHz, Galactic CR electrons involved in synchrotron emission can be assumed to be uniformly distributed over the scales of magnetic field inhomogeneities \citep[see e.g.][]{Regis11}. As a consequence, the fluctuations of synchrotron radiation emitted over a large volume and detected in total intensity radio maps directly reflect  the spectrum of magnetic fluctuations. Indeed, high dynamic range radio maps of spatially extended ISM features display fluctuations in both total \citep{Haslam82} and polarized intensity \citep{Wieringa93,Carretti09} over a wide range of spatial scales. The advantage of this method is that it relies on total intensity data that are not affected by depolarization and hence by the thermal electron density distribution. As a result, it is a powerful tool to look at spatial fluctuations of magnetic fields. An analysis of total power synchrotron fluctuations both in the Galaxy and in the nearby spiral galaxy M\,33 was recently performed by \citet{Stepanov12} in order to study magnetic turbulence.

Also, the characterisation of the diffuse synchrotron foreground at arcminute angular scales is fundamental for cosmological studies, such as e.g.\ extracting the highly red-shifted 21 cm signal from the epoch of reionisation from low-frequency observations. At these frequencies, the Galactic diffuse non-thermal radiation dominates over all other Galactic emission components (i.e.\ dust and free-free emission), thus forming a Galactic foreground screen and constituting a limiting factor for precise cosmology measurements.

The 408~MHz \citep{Haslam82} all-sky map is the most comprehensive map of Galactic diffuse synchrotron emission at about one-meter wavelength. However due to its poor angular resolution ($\sim 0.85^{\circ}$), it is not adequate for the investigation of small-scale fluctuations in the Galactic foreground emission. Moreover, the radio emission from our Galaxy at lower frequencies is still poorly known. The new generation of radio interferometers operating below $\lesssim300$~MHz will provide high-quality interferometric data at high ($\sim1\arcsec$) angular resolution, thus overcoming this present limitation.
The LOw Frequency ARray (LOFAR) \citep[see e.g. van Haarlem et al. 2012 A\&A submitted and][]{Heald11} is one of the first of the new generation radio telescopes already in operation in the frequency range $\nu\lesssim240$~MHz. Due to its large collecting area, the dense \textit{uv}-coverage at short spacings, and the high sensitivity, LOFAR can perform sensitive observations as well as wide-field and high dynamic range imaging, allowing for detailed studies of the diffuse radio continuum.

Located mostly in the second quadrant at low positive Galactic latitudes, the Fan region is a spatially extended ($\sim100^{\circ}\times30^{\circ}$), highly polarized, and synchrotron bright region. A small field in the Fan region, which contains a conspicuous circular polarized feature \citep{Bingham67,Verschuur68, Haverkorn03b}, was recently studied in detail both in total \citep{Bernardi09} and polarized \citep{Iacobelli13} intensity. We used this field to probe the relationship of Galactic magnetic field and turbulence by studying the Galactic radio synchrotron foreground. Moreover, we had the advantage that there exists a previous observation of this field with the Westerbork telescope (WSRT) at comparable frequencies \citep{Bernardi09}, which enables a comparison with the new LOFAR results.

In this paper we summarize results obtained from a 12-hour LOFAR observation of part of the Fan region. In Sect.~\ref{s:obs} we describe the data processing. In Sect.~\ref{s:imaging} we present the frequency-averaged total intensity map, displaying the amplitude fluctuations and its power spectral analysis. Then in Sect.~\ref{s:fluctuations_analysis} we derive an upper limit for the minimum size of the turbulent cells toward the Fan region and constrain the ratio of the random to total components of the Galactic magnetic field. Finally, we discuss our results in Sect.~\ref{s:discussion}, and a summary of our results and conclusions is presented in Sect.~\ref{s:conclusion}.


\section{Observations and data reduction}
\label{s:obs}

The target field was observed with LOFAR in the framework of commissioning activities. The observation was performed on 2012 January 07-08th for 12 hours (mostly during night time),  using the LOFAR high band antennas (HBAs) arranged into 57 stations. The array configuration consisted of 48 core stations (CS) and 9 remote stations (RS). The phase centre was set at right ascension $\alpha$=03:10:00.00 and declination $\delta$=$+$65:30:00.0 (J2000), and no flux calibrator was observed for the adopted single-beam observing mode. Data were recorded over the frequency range 110-174~MHz with an integration time of 2~s. This frequency range was divided into 244 subbands (each with a bandwidth of about 0.18~MHz). The longest and shortest baselines recorded correspond to $\sim81$~km and $\sim36$~m respectively, although we used baselines only up to about 12~km for better calibratability, resulting in a resolution of about $60''-80''$. radio frequency interference (RFI) flagging was done for each subband with the Default Pre-Processing Pipeline ({\tt DPPP}) using the algorithm described by \citet{Offringa10, Offringa12}.


\begin{table*}[htbp]
\centering
\caption{\label{t:data_prop} Observational properties of our LOFAR data set.}
\begin{tabular}{ll}
 & \\
\hline
\hline \\ 
Phase centre$^a$ (J2000) & $\alpha$:  03:10:00.0 ($\pm$ 0.\arcsec 2) \\
                           & $\delta$: +65:30:00.0 ($\pm$ 0.\arcsec 1) \\
Start date (UTC) & 07-Jan-2012/14:00:10.0 \\
End date (UTC) & 08-Jan-2012/02:00:10.0 \\
\hline \\ 
Frequency range & 110--174~MHz\\
Wavelength range & 172--273~cm \\
CS primary beam FWHM at 160~MHz & 4.3$^{\circ}$ \\
RS primary beam FWHM at 160~MHz & 2.8$^{\circ}$ \\
\\

\hline \\
\end{tabular} 

\end{table*}


A visual inspection of the visibilities revealed some time-dependent emission from the brightest radio sources in the sky, outside the field of view and modulated by the station beam side lobes. We find that only Cassiopeia~A and Cygnus~A cause significant spurious emission. Therefore our data reduction strategy consists of:

\begin{itemize}
\item removal of the two sources Cas~A and Cyg~A,
\item (single direction) calibration of the target field visibilities,
\item identification and removal of bad data per station,
\item self-calibration to correct for direction-dependent effects,
\item imaging.
\end{itemize}

Each data reduction step was performed using software tools of the LOFAR standard imaging pipeline \citep[for a description see e.g.][]{Pizzo10,Heald10}. Both the subtraction of A-team visibilities and the single direction calibration were performed with the BlackBoard Self-calibration ({\tt BBS}) package \citep{Pandey09}, which is based on the measurement equation \citep[see e.g.][]{Hamaker96}. In order to solve and correct for directional dependent effects we used the {\tt SAGEcal} software \citep{Kazemi11}. We now discuss each of these steps individually.


\begin{figure}
\resizebox{9cm}{!}{\includegraphics{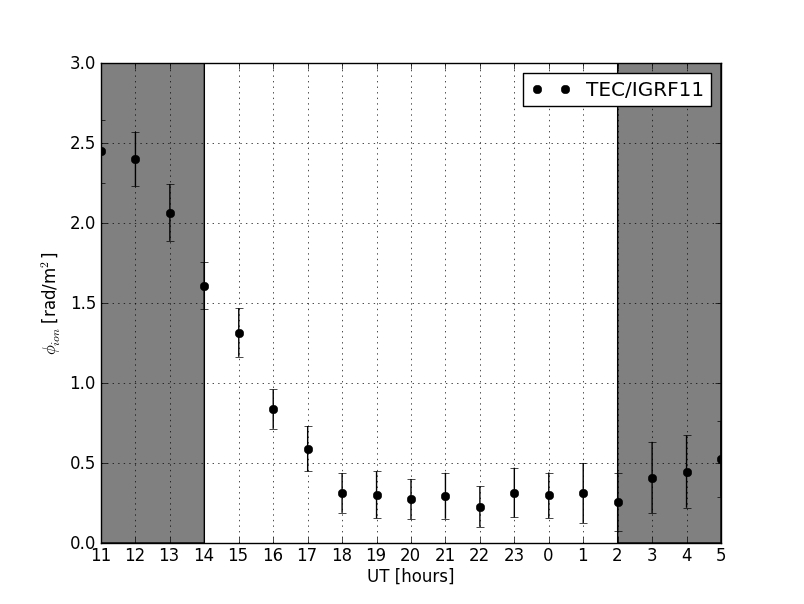}}
\caption{\label{f:ionrm} Diagnostic plot of the predicted ionospheric Faraday rotation and its time variation for this observation. Uncertainties are based only on the RMS values of the CODE global TEC maps.}
\end{figure}


\subsection{Subtraction of Cas A and Cyg A visibilities}

The removal of Cas~A and Cyg~A visibilities was done from the time-averaged data because subtraction of visibilities from full-time resolution data provided maps with a noise level about 2.5 times higher due to the lower signal-to-noise ratio (S/N) per visibility. Increasing the integration time of the gain solutions from 1~s to 20~s improved the subtraction and decreased the size of the data set. First, the direction-dependent complex gain solutions were calculated for each subband and each of the two A-team sources. Inspection of gain solutions indicated a higher impact of Cas~A than Cyg~A, likely due to its higher apparent luminosity. Then the two A-team sources were subtracted from the visibility data using their direction- dependent gain solutions. To remove residual RFI and bad data appearing after this first calibration step, another {\tt DPPP} flag step was performed on the subtracted data.

\subsection{BBS calibration}
\label{s:calibration}

The {\tt BBS} calibration needs a sky model, which was extracted from a primary beam-corrected Stokes~I map obtained in the previous WSRT observations at 150~MHz of the same field \citep{Bernardi09}. It consists of a list of clean components describing the point sources only. Furthermore, no information related to the Stokes~Q, U, or V parameters was included, and a constant spectral index of $\alpha=-0.8$ was used for all sources\footnote{We tested the impact of the assumption of a constant spectral index by comparing results obtained for a sample of five subbands. We did this by adopting a sky model with a spectral index of $\alpha=-1.0$; however, the visual inspection of the maps pointed out no differences.}. 

Low-frequency observations are affected by ionospheric propagation effects, introducing differential phase delays and Faraday rotation. Both these effects are time and direction dependent and are proportional to the total electron content (TEC) of the Earth ionosphere along the line of sight. Therefore differential Faraday rotation appears between array elements that probe different lines of sight, resulting in a phase rotation of the measured visibilities. Differential ionospheric phase rotations cause image plane effects (e.g. smearing and source deformation), while the related differential Faraday rotation affects polarization. We estimated the global TEC time variations for this observation by predicting the amount of ionospheric Faraday rotation and its time variability. To this aim we ran the {\tt ionFR} code \citep{Sotomayor13} and show in Fig.~\ref{f:ionrm} the prediction for the RM variations during the time of the observation. With the exception of the first three hours (i.e. observation during the sunset), a steady amount of ionospheric Faraday rotation of $\sim0.3$~rad~m$^{2}$ was predicted. At 146~MHz, the lowest frequency used for the next imaging step, such an RM implies a change in polarization angle of $\sim121^{\circ}$. Because we did not have any (point-like) phase reference calibrator observed, we could not directly inspect the visibility (amplitude and phase) profiles in order to search for signatures of differential ionospheric Faraday rotation. However, estimates of differential Faraday rotation in the HBA indicated phase variations of about ten degrees for baselines comparable to or larger than ours (Wucknitz, priv. comm. on LOFAR Users Forum). Moreover, we found no signal from point sources in Stokes~V maps, which indicates a very limited role of differential Faraday rotation. Therefore, we performed corrections of the visibility phases for differential phase delays using {\tt BBS} and decided to not apply Faraday rotation corrections in this analysis.

A modelled estimate of the station beam was taken into account when calculating the model visibilities. We applied the calibration solutions and corrected the data for each station beam response in the phase centre. 

\subsection{Removal of bad data}

Due to limited receiver synchronization at the time of the observation, the performance of some stations was not optimal, causing decorrelation of signals and, especially around 100~MHz, beam-shape deformation. These effects were visible in the solutions of the calibration step in these faulty stations, showing up as systematically lower gains or noisier phase patterns. This occurred in 15 stations (12~CS and 3~RS), which were subsequently flagged. Next a further flagging step was carried out and in order to minimize the beam-shape deformation effect, which is primarily present at the lower frequencies, we used only the 144 subbands at frequencies higher than 145~MHz.

\subsection{Self-calibration and imaging}
\label{s:flux_corrections}

Once the direction-independent calibration step was completed, an intermediate imaging step was done using CASA\footnote{Common Astronomy Software Applications, http://casa.nrao.edu} imager. A $16^{\circ}.7 \times 16^{\circ}.7$ total power sky map of each subband was imaged and cleaned using CASA imager with w-projection \citep{Cornwell05,Cornwell08}, but without primary beam corrections. These wide-field Stokes I maps with a resolution of $86\arcsec \times 74\arcsec$ were used to update the sky model in the next self-calibration step. Therefore, to mitigate direction-dependent errors seen in the wide-field maps we ran {\tt SAGEcal} with a solution interval of five minutes. To match our sky model, which consisted of point sources only, and to exclude extended emission from the model, we also excluded baselines shorter than 50~lambda in the creation of a clean component model from the CASA images. Finally, a flagging of the corrected data was done. 

The final imaging step of the self-calibrated dataset was performed using both the CASA and AW imagers. The sky maps for each subband were imaged with uniform weighting, allowing high resolution. Again, the CASA imager provided us with a wide-field mapping, while the AW imager \citep{Tasse13}, which is part of the LOFAR software, provided us with primary beam-corrected sky maps to be used when comparing the LOFAR and WSRT fluxes. Moreover, the AW imager is tailored to perform corrections for direction-dependent effects (e.g.\ the LOFAR beam and the ionosphere) that vary in time and frequency. Finally, each clean model sky map was convolved with a nominal Gaussian beam, and the {\tt SAGEcal} solutions were applied to the residual sky maps in order to properly restore the fluxes. 

\begin{table}[htbp]
\centering
\caption{\label{t:map_prop} Properties of individual subband (top) and frequency-averaged (bottom) Stokes~I maps.}
\begin{tabular}{ccclll}
\hline
\hline \\ 
  & {\bf CASA imager} & {\bf AW imager} \\
\hline \\
Dynamic range: & $\sim500$ & $\sim500$ \\
Rms noise$^{a}$: & 4.0 -- 3.2 & 3.8 -- 3.1 \\
Beam size: & $86\arcsec \times 74\arcsec$, PA=92$^{\circ}$ & $80\arcsec \times 70\arcsec$, PA=88$^{\circ}$ \\
Field size: & 16.7$^{\circ}\times 16.7^{\circ}$ & 10.0$^{\circ}\times 8.0^{\circ}$ \\
\\
Dynamic range: & $5.08 \times 10^3$ & $5.80 \times 10^3$ \\
Rms noise$^{b}$: & 0.40 & 0.45 \\
Beam size: & $86\arcsec \times 74\arcsec$, PA=92$^{\circ}$ & $80\arcsec \times 70\arcsec$, PA=88$^{\circ}$ \\
Field size: & 16.7$^{\circ}\times 16.7^{\circ}$ & 10.0$^{\circ}\times 8.0^{\circ}$ \\

\hline \\
\end{tabular} 
\\
{\it a)} Flux density unit is mJy~beam$^{-1}$. The values refer to the frequency range 146 -- 174~MHz.\\
{\it b)} Flux density unit is mJy~beam$^{-1}$.
\end{table}


\section{Observational results}
\label{s:imaging}

\subsection{Continuum emission maps}

\begin{figure}[b]
\resizebox{9cm}{!}{\includegraphics{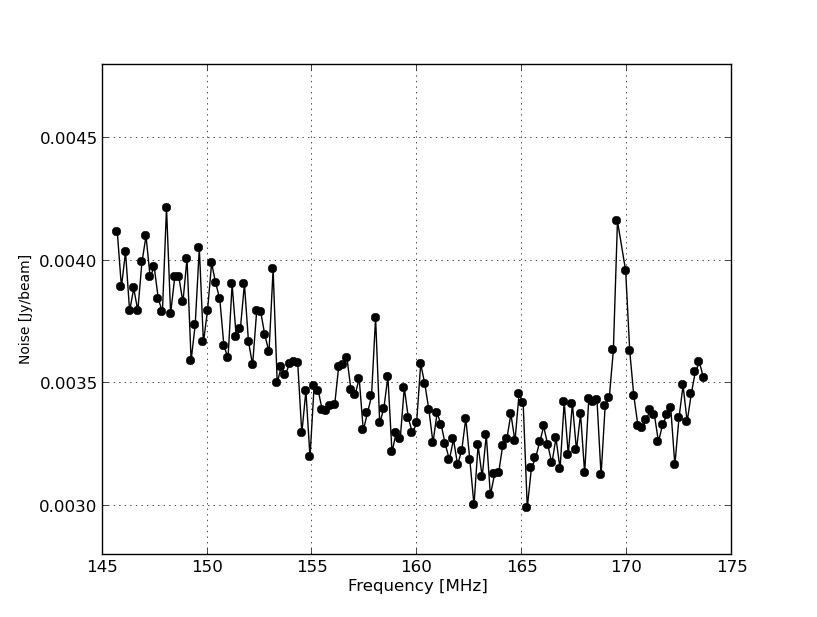}}
\caption{\label{f:noise_profile} The behaviour of the noise as a function of frequency in maps obtained with CASA imager. A prominent peaked feature in the noise level is seen around 169~MHz. A thermal noise level of about 1~mJy~beam$^{-1}$ is expected for each subband over the range 146--174~MHz for this observation.}
\end{figure}

The main features of the calibrated maps for a single subband are summarised in Tab.~\ref{t:map_prop}. Maps obtained with the CASA imager have a noise level measured out of the main beam that varies from about 4.0~mJy~beam$^{-1}$ (i.e.\ about four times the expected thermal noise level of 1~mJy~beam$^{-1}$) at $\sim$146~MHz to about 3.2~mJy~beam$^{-1}$ at $\sim$165~MHz, rising up about 3.4~mJy~beam$^{-1}$ at $\sim$174~MHz as shown in Fig.~\ref{f:noise_profile}. An evident spike is found around 169~MHz, and four related subbands of the CASA imaging were discarded. Maps of each subband were inspected visually after the imaging step with AW imager; 17 primary beam-corrected maps displaying an extended pattern of artifacts propagating from the source 4C+63.05 at the south-west edge of the field had to be discarded.


\begin{figure*}[htbp]
\resizebox{19cm}{!}{\includegraphics{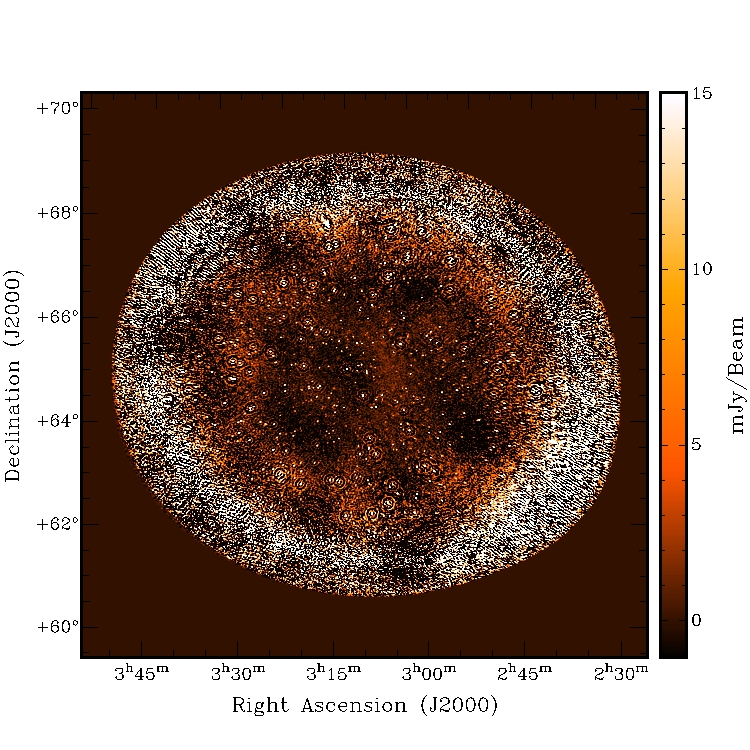}}
\caption{\label{f:stokesI_AWmap} Frequency-averaged Stokes~I map of the Fan region field, as obtained with the AW imager with a resolution of $80'' \times 70''$.}
\end{figure*}

\begin{figure}[htbp]
\resizebox{9cm}{!}{\includegraphics{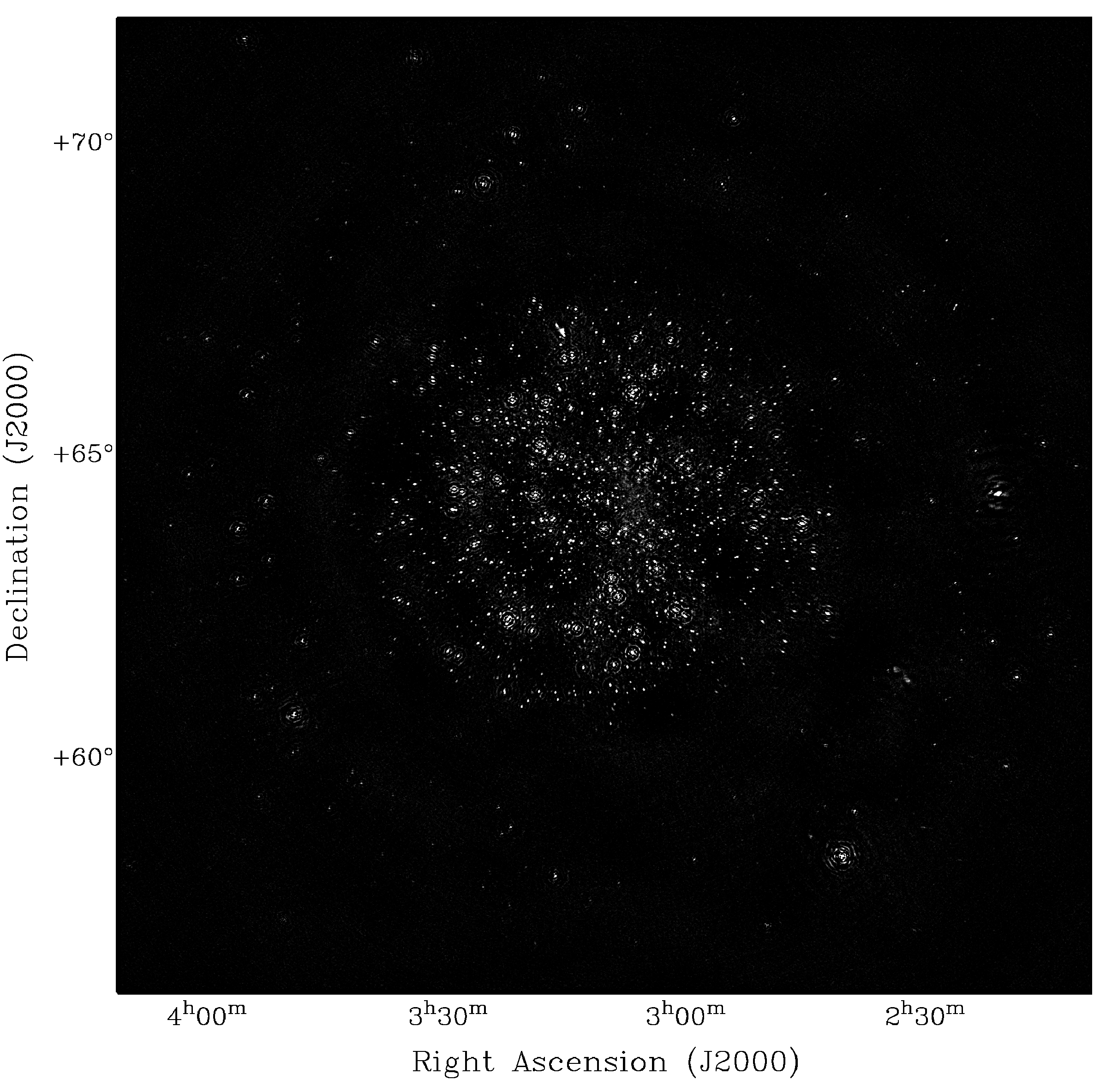}}
\caption{\label{f:stokesI_CASAmap} Frequency-averaged Stokes~I map of the Fan region field, as obtained with the CASA imager with a resolution of $86'' \times 74''$. The bright sources out of the main beam showing artifacts are 4C+58.08, 4C+72.06, and 4C+64.02.}
\end{figure}

In a single subband map of total intensity, many extragalactic point sources are visible as well as artifacts around bright sources, but no Galactic diffuse emission is detected. In order to increase the S/N ratio, the individual subbands were combined into one frequency-averaged map. 
The LOFAR main beam, frequency-averaged map after the imaging step with AW imager, which is primary beam corrected, is given in Fig.~\ref{f:stokesI_AWmap}.  Fig.~\ref{f:stokesI_CASAmap} depicts the full bandwidth-averaged map covering $16^{\circ}.7 \times 16^{\circ}.7$, which has a measured noise level out of the main beam of $\sim0.4$~mJy~beam$^{-1}$ and a dynamic range of 5080. The imaging step with the AW imager after the self-calibration results in a slightly higher noise level of $\sim0.45$~mJy~beam$^{-1}$ and a slightly higher dynamic range $\sim5800$. The resulting maps are confusion dominated toward its centre; indeed at 160~MHz and with a beam size of about 1$\arcmin$, the expected confusion noise level is about 1~mJy~beam$^{-1}$ \citep{Brown11}. 


The CASA imaged map (Fig.~\ref{f:stokesI_CASAmap}) clearly shows hundreds of point sources and a few extended extragalactic sources within the primary beam as well as a significant number of extragalactic unresolved sources out of the primary beam. Furthermore, artifacts are evident around bright sources spread within the imaged field, indicating a limited accuracy of calibration. The brightest sources in the imaged field are 4C+58.08, 4C+72.06, and 4C+64.02 with fluxes at 178~MHz of about 19.9, 9.6, and 7.6~Jy respectively. All these sources are located out of the main beam, but only 4C+58.08 and 4C+64.02 show evident artifacts. This is likely because the sky model treats these sources as single point sources, while their structure is partially resolved at the adopted angular resolution.

The primary beam-corrected Stokes~I map imaged with AW imager also displays hundreds of point sources as well as artifacts around bright sources, but now the noise dominates towards the edges. Intriguingly, we detect diffuse and faint continuum in both frequency-averaged maps toward the Fan region, at a level of about 3~mJy~beam$^{-1}$. The complex spatial morphology agrees with Stokes~I structure seen in the WSRT map at lower resolution \citep[see Fig.5 of][]{Bernardi09}. In what follows we focus on this faint, very extended, Galactic emission. Since the detected diffuse emission is relevant for both cosmological and foreground studies, we describe its spatial properties statistically through its angular power spectrum. 


\subsection{Comparing LOFAR with WSRT data}
\label{s:comp}

\begin{figure}
\resizebox{9cm}{!}{\includegraphics{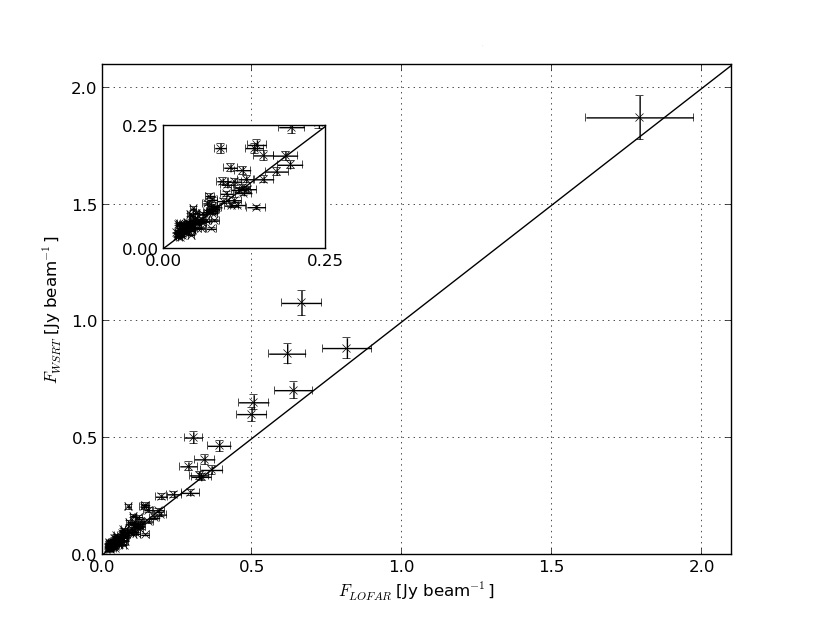}}
\caption{\label{f:LOFAR_WSRT_fluxes} Comparison between the LOFAR fluxes rescaled to 150~MHz and the WSRT fluxes at 150 MHz of point sources detected above a threshold of 20~mJy~beam$^{-1}$ within a $3^{\circ}\times 3^{\circ}$ box centred at the phase centre. The reference flux ratio of unity is indicated by the solid line.}
\end{figure}

\begin{figure}
\resizebox{9cm}{!}{\includegraphics{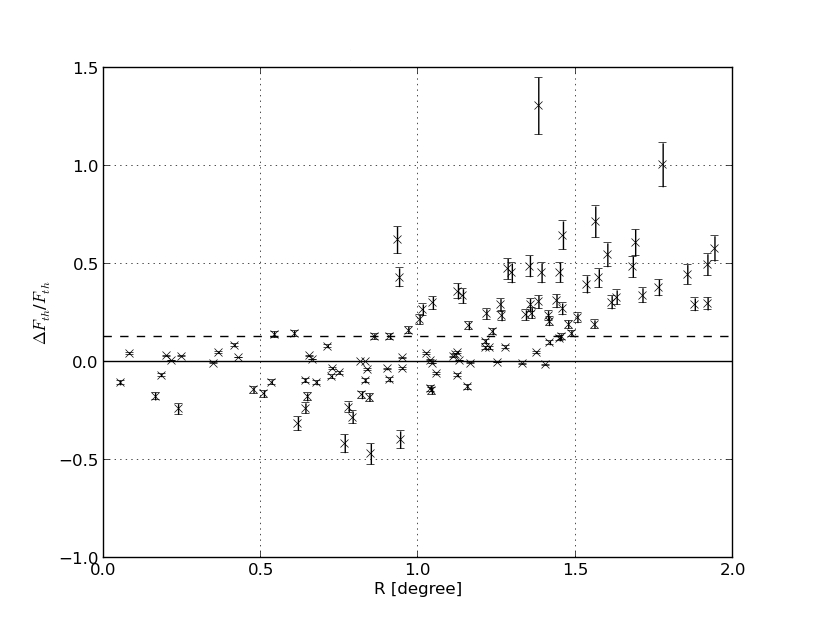}}
\caption{\label{f:frac_flux_freq_corr_20mJy} The normalized peak flux differences between point sources in the LOFAR and WSRT observations as a function of radial distance. The LOFAR peak fluxes were rescaled to 150~MHz and both maps were corrected for primary beam attenuation. The mean value (dashed line) of these fractional variations is $(\Delta F / F)=(0.128\pm0.014)$, indicating the presence of a bias.}
\end{figure}

We test the quality of the LOFAR flux calibration by comparing the point sources in the frequency-averaged Stokes~I map to sources detected at this frequency with the WSRT \citep{Bernardi09}. We select sources stronger than 20~mJy~beam$^{-1}$ in a $3^{\circ}\times 3^{\circ}$ region centred at the phase centre in the LOFAR map. We rescale the LOFAR fluxes measured at a reference frequency of 160~MHz to the WSRT reference frequency of 150~MHz, using a constant spectral index of $\alpha=-0.8$. The error in the WSRT flux density is 5\% \citep{Bernardi09} and the LOFAR flux uncertainty was assumed at a level of 10\%. The fluxes of point sources measured in the LOFAR and WSRT maps are compared in Fig.~\ref{f:LOFAR_WSRT_fluxes}. The LOFAR fluxes of this sample of sources are mostly consistent with WSRT ones. Small deviations from the reference flux ratio may reflect either residuals of calibration or a different spectral behaviour. However, the differences in LOFAR and WSRT flux of these point sources seem to be systematic in position. 
We compare the LOFAR peak fluxes above a threshold of 20~mJy~beam$^{-1}$ within a $3^{\circ}\times3^{\circ}$ box centred at the phase centre rescaled to the WSRT reference frequency to those from the WSRT primary beam-corrected map. The corresponding peak fluxes are used as a reference for the calculation of the relative flux difference $\frac{\Delta F}{F}$:
\begin{equation} 
\frac{\Delta F}{F} = \frac{F_{WSRT}-F_{LOFAR}}{F_{LOFAR}} \, .
\end{equation} 
The relative flux difference as a function of the radial distance from the field centre is shown in Fig.~\ref{f:frac_flux_freq_corr_20mJy}. Out to a radius of about one degree from the phase centre, the LOFAR and WSRT fluxes agree well (slightly worse for the weakest sources) and a flat $\Delta F/F$ profile is seen, while at larger radii the LOFAR fluxes are increasingly lower than the WSRT fluxes. We explain this systematic effect in the image plane as due to the combination of core and remote stations having different beams with a size of about 4.6 and 3.0 degrees FWHM at 150~MHz respectively. Therefore, out of a region with a radius of about 1.5 degrees, the resolution is expected to decrease (by about a factor 4) because of the smaller contribution to the visibilities of the remote stations, thus affecting the measured peak fluxes. In the following, we use the inner ($3\times3$~degrees) part of the field of view only to mitigate this systematic effect. Also, an evident scatter of data points is found over the entire range of radial distances, which may indicate a limited accuracy of the LOFAR beam model (e.g.\ a non-negligible azimuthal dependence), but we note that the errors in the WSRT beam model, which is poorly known at such low frequencies, are also present in the comparison. 


\subsection{Power spectral analysis}
\label{s:power_analysis}

To perform the angular power spectral analysis two approaches are feasible, namely working in the image plane or directly in the visibilities$'$ \textit{uv}-space. The first allows calculation of the angular power spectrum of a selected sky region, thus permitting the contributions of different astrophysical sources to be separated from the bulk of detected power; however, it is affected by systematics due to the imaging step. The latter provides a proper errors estimate and investigation of data quality and systematics effects but does not allow  contributions towards different directions in the sky to be distinguished. In this study, both these issues are relevant and the approaches are complementary. 

In order to evaluate the distribution of detected power in the \textit{uv}-plane we consider the calibrated, residual visibilities. To convert the power to squared temperature brightness, we need to estimate the size of the main beam seen at station level. Indeed, the sensitivity in the plane of the sky of a receiving LOFAR station is a function of the observing frequency and the size of the station, and LOFAR has stations of two types and sizes, the CS and RS stations respectively. Therefore a main beam with different angular sizes is formed by core-core core-remote and remote-remote baselines, and we correct for this effect by assuming a cylindrical approximation for the beam shape. The power at angular scales we are interested is mainly detected by CS. Thus we select the visibilities from CS-CS baselines only as a function of the \textit{uv}-distance with a maximum \textit{uv}-range of 10~k$\lambda$, calculate the Stokes~I parameter and finally the power spectrum.

\begin{figure*}
\centering
\resizebox{1.0\columnwidth}{!}{\includegraphics{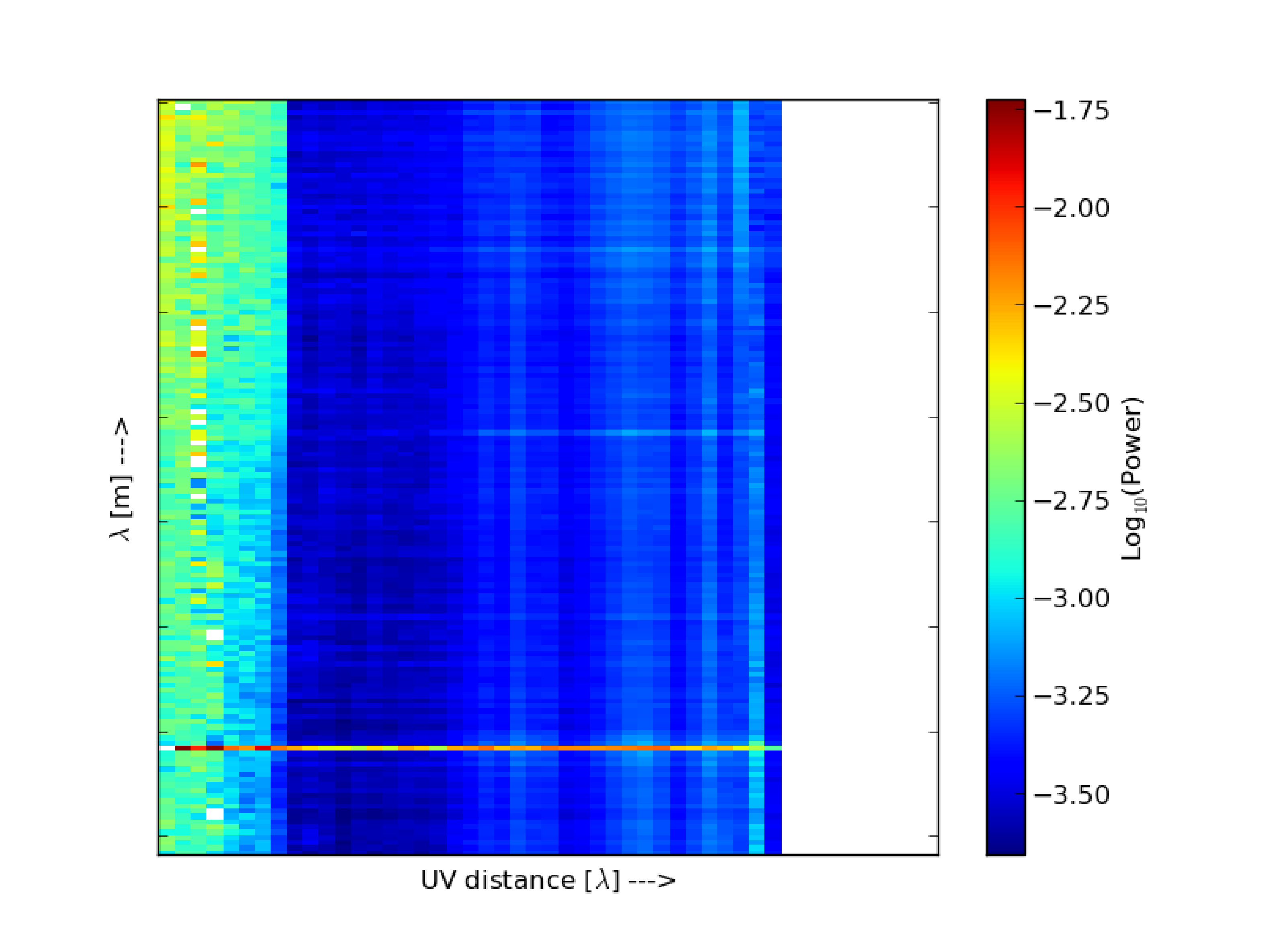}}
\resizebox{1.0\columnwidth}{!}{\includegraphics{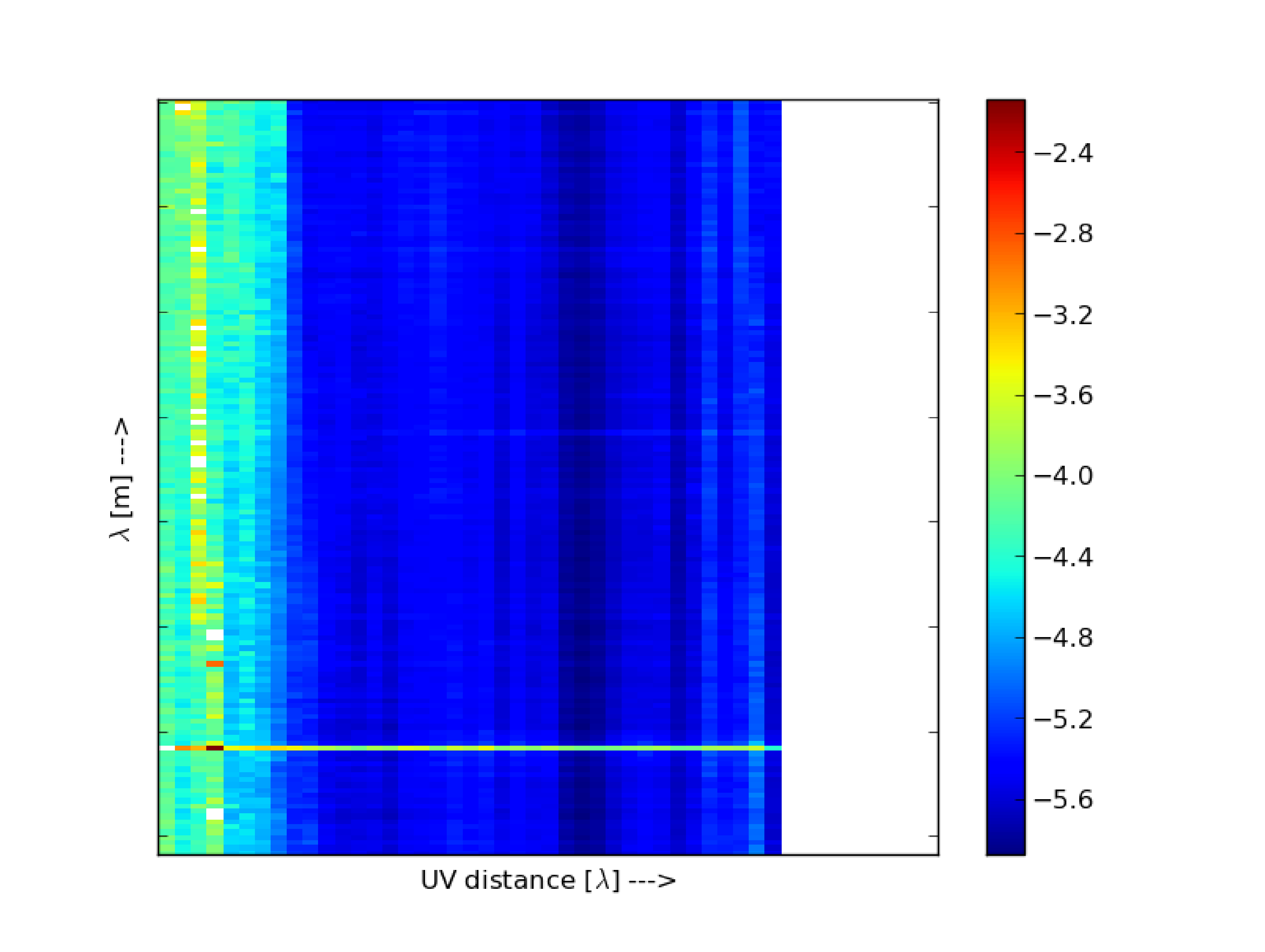}}
\caption{\label{f:uv_powspec} Distribution of the power (left panel) and its error (right panel) as a function of the \textit{uv}-distance (up to 10~k$\lambda$) and frequency, detected by CS stations only. The row showing a systematic excess of power over the entire \textit{uv}-distance range corresponds to the corrupted and discarded SB\,233.}
\end{figure*}

As a result, we obtain the multi-frequency angular power spectrum shown in Fig.~\ref{f:uv_powspec}, where an evident excess of power at short baselines (i.e. at large angular scales) is displayed. Also, a frequency dependence of this large scale emission is seen, the larger amount of power being towards long wavelengths. The systematic excess of power over the entire range of \textit{uv}-distances indicates the presence of instrumental effects corrupting the data, and therefore we exclude SB~233.

The angular total power detected by LOFAR from the observed target field is the sum of several contributions. The diffuse Galactic foreground (which is not modelled),
 consists of the synchrotron fluctuations due to MHD turbulence spread across the field of view, the presence of an extended and nearby \citep{Iacobelli13} Galactic object close to the phase reference, the extended W3/4/5 \ion{H}{ii} region complex, and the Galactic plane emission towards the lower west edge of the observed field at a radial distance of $\sim5.6$ and $\sim6.4$ degree from the phase reference. Also, at sub-degree scales, the spiral galaxy IC\,342 and the giant double lobe radio galaxy WBN\,0313+683, which are located at $\sim4.4$ and $\sim3.7$ degrees from the phase reference respectively, produce power excess. 

The only way to perform spatial selection in the \textit{uv}-domain is to tune the station field of view by selecting a proper frequency range. In this way we can minimize power contributions due to the Galactic plane and the extended W3/4/5 \ion{H}{ii} region complex, the price being the use of only a fraction of the data.


\begin{figure}[htbp]
\resizebox{9cm}{!}{\includegraphics{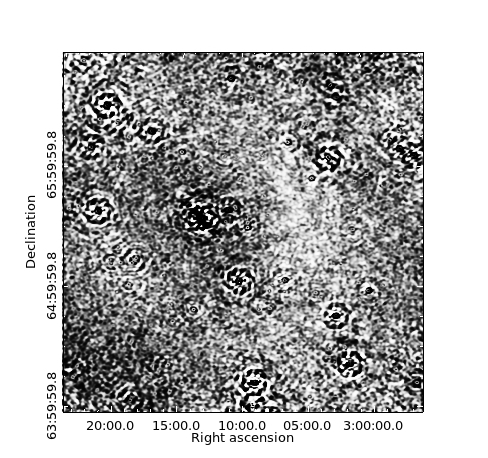}}
\caption{\label{f:residual_map} The inner $3^{\circ}\times3^{\circ}$ region of the primary beam-corrected Stokes~I map with point sources subtracted.}
\end{figure}

To avoid this drawback and discard the unwanted power contributions, we use the prescription by \citet{Bernardi09} to calculate the power spectrum. However, instead of identifying the point sources by making sky images with only the long baselines and subtracting these directly from the visibilities, as Bernardi et al.\ did, we identify and extract point sources from the frequency-averaged total intensity map down to $\lesssim5$~mJy~beam$^{-1}$ using the {\tt PyBDSM} source extraction software\footnote{http://tinyurl.com/PyBDSM-doc}. We obtain the residual image shown in Fig.~\ref{f:residual_map}, where an extended pattern of fluctuations is seen, along with evident artifacts around bright sources; only very faint sources are left. 


From this residual map we calculate the power spectrum as in \citet{Seljak97} and \citet{Bernardi09} over a region of $3^{\circ}\times3^{\circ}$~degrees centred on the field centre:
\begin{eqnarray}
C^{I}_{\ell} = \left\{ \frac{\Omega}{N_{\ell}} \sum_{\bf \ell} I({\vec\ell}) I^*({\vec\ell}) - \frac{\Omega (\sigma^I_{\rm{noise},\ell})^2}{N_b} \right\} b^{-2}({\vec\ell}) \:,
\label{pow_spec_def}
\end{eqnarray}

where $I$ indicates the Fourier transform of the total intensity, $\ell$ is the multipole (i.e. $\ell={180}/{\theta}$ where $\theta$ is the angular scale in degrees), $\Omega$ is the solid angle in radians, ${N_{\ell}}$ is the number of Fourier modes around a certain $\bf{\ell}$ value, $\sigma_{\rm{noise},\ell}$ is the noise per multipole, and $N_b$ is the number of independent synthesized beams in the map. In the case of negligible calibration errors, the factor $b^{2}(\ell)$ is the power spectrum of the window function \citep{Tegmark97}, which in the case of interferometric images corresponds to the power spectrum of the dirty beam. However, this is not the case for the LOFAR data, as indicated by the presence of artifacts in the image plane. Therefore, taking into account only the dirty beam provides just approximate corrections to the power spectrum. Thus, instead of being able to measure angular size scales down to the synthesized beam size from the interferometry, we are limited to the largest size scale ($\sim8\arcmin$) of the imaging artifacts shown in Fig.~\ref{f:residual_map}.

\begin{figure}
\resizebox{9cm}{!}{\includegraphics{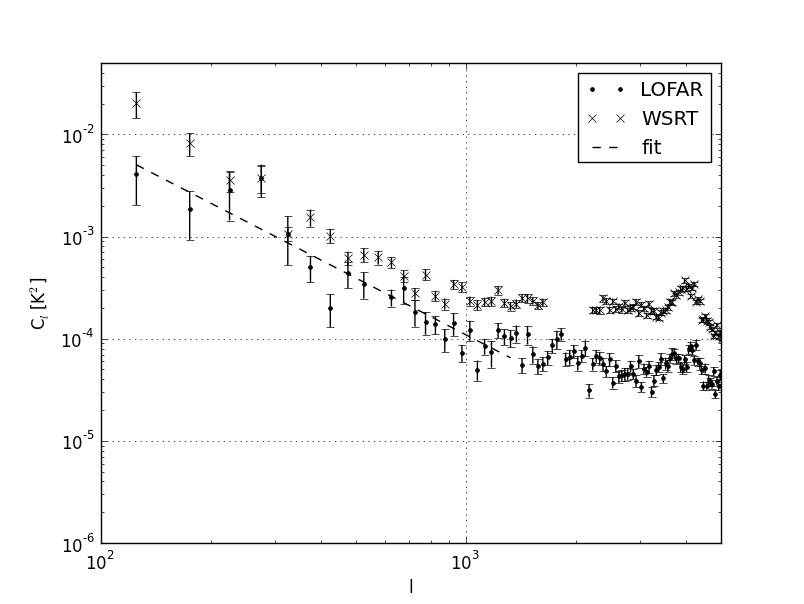}}
\caption{\label{f:pow_spec} Power spectra of total intensity from the LOFAR (dots) and WSRT (crosses) observations. The error bars indicate statistical errors at $1\sigma$. The fitted power law (dashed line) with a spectral index $\alpha=-1.84\pm 0.19$ for $\ell\in[100,1300]$ is also shown.}
\end{figure}

The largest angular scale of emission measured by an interferometer, i.e.\ the smallest multipole $\ell_{min}$, is fixed by the shortest (u,v) spacings of the interferometer. For LOFAR $\ell_{min} \gtrsim 50$. In principle, a constraint on the smallest scale suitable for the investigation of the data is given by the size of the point spread function (PSF), which attenuates the angular power; at LOFAR angular resolution ($\sim1\arcmin$) attenuation would be negligible up to $\ell_{max} \lesssim 3700$. As noted above, because of the artifacts affecting the image plane due to non-negligible calibration errors we practically limit our investigation up to $\sim8\arcmin$, corresponding to a multipole $\ell_{max} \lesssim 1350$. We used a least square method to fit a power law ($C_{\ell} \propto \ell^{\alpha}$) to the angular power spectrum, giving a spectral index $\alpha=-1.84\pm 0.19$ for $\ell\in[100,1300]$ in agreement within 2 sigma with the previous slope estimate from WSRT data \citep{Bernardi09}. 
The power spectrum down to scales of about $2\arcmin$ is shown in Fig.~\ref{f:pow_spec}. The uncertainties were calculated as the standard deviation of the signal within one multipole bin. At low $\ell$ the power spectrum shows a power law behaviour, while at high multipoles (i.e. $\ell>10^3$) a flat power spectrum is seen. \citet{Bernardi09} interpreted this power law as the large-scale foreground emission from the Galaxy toward the Fan region and the flattening of the spectrum as the rms confusion noise ($\sigma_{c}$) due to the point source contamination. As in \citet{Brown11}, we can estimate $\sigma_{c}$ as
\begin{equation}
\frac{\sigma_{c}}{\mbox{mJy~beam}^{-1}} \approx 0.2\times\left(\frac{\nu}{\mbox{GHz}}\right)^{-0.7}\times\left(\frac{\theta}{\mbox{arcmin}}\right)^{2} \:,
\end{equation} 
where $\nu$ is the reference frequency and $\theta$ is the FWHM of the Gaussian beam. The confusion noise level for the LOFAR data is $\sigma_{c}\sim0.85$~mJy~beam$^{-1}$, which corresponds to a power of $C_{\ell}^c \sim 4\times10^{-5}$~K$^{2}$. 

For comparison, the WSRT power spectrum obtained by \citet{Bernardi09} is overplotted in Fig.~\ref{f:pow_spec}. The shapes of the power spectra from the LOFAR and WSRT data match very well, indicating that the diffuse emission in the LOFAR data probes the same total intensity fluctuations as the WSRT. However, the amplitude of the LOFAR data is a factor 3 or so lower, a plausible reason being the applied weighting scheme. In order to clarify this point we performed an imaging step with a natural weighting scheme, thus not changing the power distribution at different angular scales. Now we obtain an amount of power in LOFAR profile that was consistent with the WSRT one, with no change of the spectral shape in the multipoles$'$ range of interest (i.e.\ $\ell \lesssim 1300$). However, the beam is about three times worse and more diffuse emission is recovered, thus implying a less accurate source subtraction step with {\tt PyBDSM}. Because of the higher rms confusion noise, a lower maximum $\ell$ value characterizing the power law is obtained and we decided to use the uniform weighting scheme. 
 


\section{Turbulence of the diffuse Galactic foreground}
\label{s:fluctuations_analysis}

In the following, we use power spectra fluctuations to constrain the outer scale of turbulence ($L_{out}$) of the Galactic synchrotron foreground (Sec.~\ref{s:lout}). In Sec.~\ref{s:bratio}, we use this information to deduce the ratio of the regular to random field strength ($B_{o}/B_{r}$) as a function of $L_{out}$. 

\subsection{The outer scale of fluctuations}
\label{s:lout}

Magnetic turbulence in the Galactic disk and the halo dictates the power spectral behaviour of synchrotron intensity. Based on earlier results of \citet[][]{LazarianShutenkov90}, \citet[][]{ChoLazarian02} modelled the synchrotron emissivity in two different regimes: where the angle between the lines of sight is so small that they travel mostly through the same turbulent cells (i.e.\ $\theta < L_{out}/L_{max}$, with $L_{max}$ being the path length to the farthest turbulent cells), and where the angle between the lines of sight is large, so that these lines of sight mostly probe independent cells. They conclude that on the small scales the synchrotron power spectrum should show the same (Kolmogorov) slope as the magnetic field power spectrum. On the large scales, however, the slope will be shallower and a function of the Galactic latitude. The latter effect occurs because at higher Galactic latitudes there will be more emissivity near the disk than further away in the halo. The scale at which the power spectrum transits from Kolmogorov to a shallower slope is the critical scale $\ell_{cr} \sim \pi L_{max} / L_{out}$.

For a characteristic scale height of the Galactic synchrotron emission of $H_{sync}=1.0\pm0.4$~kpc \citep[see e.g.][]{Page07}, the path length $L_{max}$ through the Galactic synchrotron layer is $L_{max}=H_{sync}/\sin b$, with $b$ the Galactic latitude. At the latitude $b=+7^{\circ}$ of the Fan region, the distance up to the boundary of the probed volume is $L_{max}=8.2\pm1.6$~kpc. This path length indicates an average emissivity of $\epsilon_{b=7^{\circ}}=5.5\pm1.1$~K/kpc, in agreement with the synchrotron emissivity of $\sim7$~K/kpc at 408~MHz in the solar neighborhood \citep{Beuermann85}. 
The LOFAR power spectrum corresponds to the shallow (large angular scale) regime of the model by \citet{ChoLazarian02}, which means that the critical multipole $\ell_{cr}$ is a multipole larger than the higher multipole of the power law before the spectrum flattens to noise. The higher multipole of the power law is $\ell\sim 1300\pm300$, indicating an outer scale of turbulence of $L_{out}\lesssim20\pm6$~pc.

\subsection{Constrain $B_o/B_r$ from $L_{out}$}
\label{s:bratio}

The importance of statistical investigations of the Galactic MHD turbulence and its properties has been recently exploited by \citet{Lazarian12}, providing an accurate and quantitative description of the synchrotron fluctuations for an arbitrary cosmic ray spectral index. However, because the artefacts affecting the accuracy of the data and the calculated power spectrum do not allow us to aim with such a precision, we adopt the earlier model of \citet{Eilek89a,Eilek89b} with a fixed cosmic ray spectral index of about three. The effects of MHD turbulence in subsonic and transonic regimes on the total and polarized intensity of an extended radio source were explored by \citet{Eilek89a,Eilek89b} under the assumptions that the characteristic outer scale ($L_{out}$) is much smaller than the source size ($L_{max}$) and the fluctuations obey a Gaussian statistics. This author shows how (strong) MHD turbulence produces detectable fluctuations in total intensity and how the mean ($\langle I \rangle$) and variance ($\sigma_{I}$) of the total intensity of an extended synchrotron source can be interpreted in terms of the total intensity source function 
\begin{equation}
S_I = S_0 \left(\frac{B_{\perp}}{B_o}\right)^{\frac{\gamma+1}{2}}
\end{equation}
and its standard deviation $\sigma_{SI}$, where $B_{\perp}$ is the magnetic field component perpendicular to the line of sight, $B_0$ is the ordered field component, and $\gamma$ is the spectral index of the electron energy distribution $N(E) = N_0 E^{-\gamma}$. Then the fractional source-function variance ($(\sigma^{2}_{S_{I}})^{1/2}/S_{I}$) is given by 
\begin{equation} 
\frac{(\sigma^{2}_{SI})^{1/2}}{S_{I}} \simeq \sqrt{\frac{L_{max}}{L_{out}}} \frac{\sigma_{I}}{\langle I\rangle} \:,
\label{eq:eilek89a}
\end{equation} 
\citep{Eilek89a}. Because variations of $S_{I}$ reflect variations in the random magnetic field, the ratio of source-function variance and mean points to an estimate of the ratio of the random to ordered magnetic field strengths within the extended source. For subsonic turbulence, the fluctuations in synchrotron emission are likely predominantly caused by magnetic field fluctuations, with only slightly varying relativistic electron and positron densities, so that $(\sigma^{2}_{SI})^{1/2}/\langle S_I\rangle \approx B_r^2/(B_r^2+B_o^2)$. However, for transonic turbulence, density fluctuations will also be important and the source function behaviour can be approximately represented as $S\propto B^4$ \citep{Eilek89b}, so that $(\sigma^{2}_{SI})^{1/2}/\langle S_I\rangle \approx B_r^4/(B_r^2+B_o^2)^2$. A major dependence by density fluctuations is also expected for supersonic turbulence, but this case is not treated by \citet{Eilek89b}. 

We compute the ratio of random to ordered magnetic field $B_o/B_r$ for our LOFAR data set using the prescriptions above. Since this is an interferometric measurement, short spacing information is missing and we cannot estimate $\langle I \rangle$ from our data. Instead, we obtain $\langle I \rangle$ from the absolute flux calibrated all-sky map at 408~MHz \citep{Haslam82}: the Stokes~I brightness temperature at 408~MHz is about $T_{408}^{I}=45.0\pm4.5$~K around $(l,b)\approx(137.0^{\circ},+7.0^{\circ})$. The frequency dependence of the spectral index for the synchrotron brightness temperature has been investigated by several authors, e.g. \citet[][]{de Oliveira-Costa08} and recently \citet[][]{Kogut12}. For a spectral index of $\beta=-2.64\pm0.03$ \citep{Kogut12}, the corresponding sky temperature at 150~MHz is $T_{150}^{I}=632\pm32$~K. Next the isotropic extragalactic background component is subtracted. We scale the value of about 28~K at 178~MHz of \citet{Turtle62} to 150~MHz, obtaining a contribution of about 45~K. Therefore the final sky temperature at 150~MHz is $T_{150}^{I}=587\pm30$~K, which is in agreement with the sky temperature of about 600~K around $(l,b)\approx(137.0^{\circ},+7.0^{\circ})$ of the \citet{LandeckerWielebinski70} survey at 150~MHz (Reich priv. comm.). From the residual map after source subtraction, we estimate the Stokes~I variance from the FWHM to be about $2.9$~mJy~beam$^{-1}$, which corresponds to about 22~K. Suitably scaled at 150~MHz, this value corresponds to a Stokes~I variance of 25~K. 

Observational studies of turbulence in the warm ionised medium indicate a transonic \citep{Hill08, Gaensler11, Burkhart12} regime. Rewriting Eq.~\ref{eq:eilek89a} for the transonic case gives 
\begin{equation}
\frac{B_o}{B_r} = (A^{1/2}-1)^{1/2} \;\;\;\;\mbox{where}\;\;\; A = \frac {\langle I\rangle}{\sigma_I}\left(\frac{L_{out}}{L_{max}}\right)^{1/2}
\end{equation}
For a turbulent outer cell size $L_{out}\lesssim20$~pc, the ratio of magnetic field strengths $B_o/B_r \gtrsim 0.3$.

\section{Discussion}
\label{s:discussion}

In the following we interpret the detected total power fluctuations as being due to turbulence in the magnetic field. However, we note the Fan region to be peculiar, characterised by a high polarization degree, whose origin is still debated, and implying that the regular component of the magnetic field is dominant over the turbulent one. Therefore the comparison of total power and polarisation is a major point to associate the Stokes~I fluctuations with turbulence in the magnetic field. A simple explanation is obtained by considering the different spatial depths probed by the low-frequency Stokes~I and PI emission. At low frequencies, polarisation data are constrained by the polarization horizon ($d_{h} \lesssim 1$~kpc), while the total power data can probe a larger volume and the Stokes~I fluctuations may probe different conditions along the line of sight. The polarised emission would originate in a nearby volume with a dominant ordered field, in agreement with the model of \citet{Wilkinson&Smith74}, while the total power fluctuations would arise in a farther and disordered region. \\

Many theoretical and numerical simulation results suggest the MHD turbulence in the ISM to be Alfv\'{e}nic, with an angular power spectrum matching the Kolmogorov one (i.e. a spectral index $\alpha_{K}\approx-3.7$). Observational results from investigations of H$\alpha$ emission \citep[see e.g.][]{ChepurnovLazarian10} support the Kolmogorov spectrum for the electron density fluctuations, in agreement with a weakly compressible and low Mach number turbulence. However, spectral indices steeper than $\alpha_{K}$ are inferred from the velocity fluctuations in the neutral cold phase of the ISM \citep{Chepurnov10}, indicating a high Mach number turbulence.
The analysis of synchrotron fluctuations deals with magnetic fluctuations, and  previous studies of the angular power spectrum of the Galactic radio diffuse synchrotron emission \citep[see e.g.][]{Giardino01,LaPorta08} have shown it can be fitted by a power law, $C_{l}\propto l^{\alpha}$, with $\alpha\sim[-3.0,-2.5]$ and $l\lesssim200$. Moreover, strong local variations in the index exist. However, the lower values of $\alpha$ typically correspond to the higher latitudes in both total intensity and polarized intensity \citep{Haverkorn03a}, indicating a latitude dependence of Galactic turbulence. Indeed, over the range $100\lesssim \ell \lesssim800$ \citet{Baccigalupi01} find that regions at low and medium Galactic latitudes show total intensity fluctuations with slopes displaying large variations (from $\simeq-0.8$ to $\simeq-2$), and steeper slopes corresponding to regions where diffuse emission dominates. The synchrotron spectral index ($\alpha\approx-1.84$) that we find in the Fan region is smaller than the sky-averaged spectral index of about $-2.4$ \citep{Giardino01} but consistent with the range of slopes found by \citet{Baccigalupi01}. The origin of such spatial variations of the angular power spectral features of the Galactic diffuse emission was addressed by \citet{ChoLazarian02,ChoLazarian10} in the framework of MHD turbulence with a Kolmogorov spectrum, as a result of the inhomogeneous distribution of synchrotron emissivity along the line of sight arising from the structure of the Galactic disk and the halo. 
We note that the aforementioned framework is consistent with the \citet{Goldreich&Sridhar95} theory of Alfv\'{e}nic turbulence for both a weakly compressible and a supersonic compressible \citep{ChoLazarian03} medium because a Kolmogorov spectrum is also predicted by the \citet{Goldreich&Sridhar95} model of turbulence.
To our knowledge, earlier angular power spectra of the Galactic radio diffuse synchrotron fluctuations at sub-degree angular scales and at a frequency of 150~MHz are only provided by \citet{Bernardi09}, \citet{Bernardi10}, and \citet{Ghosh12}, obtained with the WSRT and Giant Metrewave Radio Telescope (GMRT) respectively. The LOFAR and WSRT \citep{Bernardi09} studies both discuss the Fan region at low Galactic latitude $b\approx+7.0^{\circ}$, while \citet{Ghosh12} focused on two sky patches at Galactic latitudes of $b\approx+25^{\circ}$ and $b\approx+30^{\circ}$ respectively. However, due to baselines corruptions the quality of the power spectra is not as good as for the Fan region field at lower Galactic latitude. Because the power law behaviour is not well sampled, we decide to not use these WSRT data for our analysis. \citet{Ghosh12} discuss four fields of view in their paper, but only their FIELD~I at $(l,b) = (151.80^{\circ}, 13.89^{\circ})$ was used for foreground analysis because it provided the best sensitivity and point source subtraction. At $b=+7^{\circ}$ and $b=+14^{\circ}$, the distances up to the boundary of the Galactic synchrotron disk as defined above are $L_{max}=8.2\pm1.6$~kpc and $L_{max}=4.1\pm0.8$~kpc respectively. These path lengths indicate an average emissivity of $\epsilon_{b=7^{\circ}}=5.5\pm1.1$~K/kpc and $\epsilon_{b=14^{\circ}}=9.7\pm2.0$~K/kpc, which match with the synchrotron emissivity of $\sim7$~K/kpc at 408~MHz and at the solar position \citep{Beuermann85}.

The WSRT power spectrum by \citet{Bernardi09} does not show any break up to $\ell=900$; therefore $\ell_{cr}>900$ and $L_{out}\lesssim29\pm6$~pc. \citet{Ghosh12} derived an angular power spectrum in the GMRT field that does not show any break up to $\ell=800$, thus implying $L_{out}\lesssim16\pm3$~pc. We summarize the upper limits for $L_{out}$ derived from the available data in Tab.~\ref{t:lout_powspec}.


\begin{table*}[htbp]
\centering
\caption{\label{t:lout_powspec} Summary of upper limits for $L_{out}$ and $B_o/B_r$ obtained from available data at 150~MHz.}
\begin{tabular}{llccccccc}
\hline
\hline \\
Reference & Telescope&($l,b$) coordinates&$L_{max}$~[kpc]&$\ell_{cr}$&$L_{out}$~[pc] & $\langle I\rangle$~[K] & $\sigma_I$~[K] & $B_0/B_r$ \\
\hline \\
This paper & LOFAR & $137.00^{\circ},7.00^{\circ}$ & $8.2\pm1.6$ & $>1300$ & $<20\pm6$ & $587\pm30$ & $>25$ & $>0.28$\\
\citet{Bernardi09} & WSRT & $137.00^{\circ},7.00^{\circ}$ & $8.2\pm1.6$ & $>1000$ & $<29\pm6$ & $587\pm30$ & $>30.5$ & $>0.26$\\
\citet{Ghosh12} & GMRT & $151.80^{\circ},13.89^{\circ}$ & $4.1\pm0.8$ & $>800$ & $<16\pm3$ & $516\pm26$ & $>20$ & $>0.51$ \\
\hline \\
\end{tabular} 
\end{table*}

\begin{figure}
\resizebox{9cm}{!}{\includegraphics{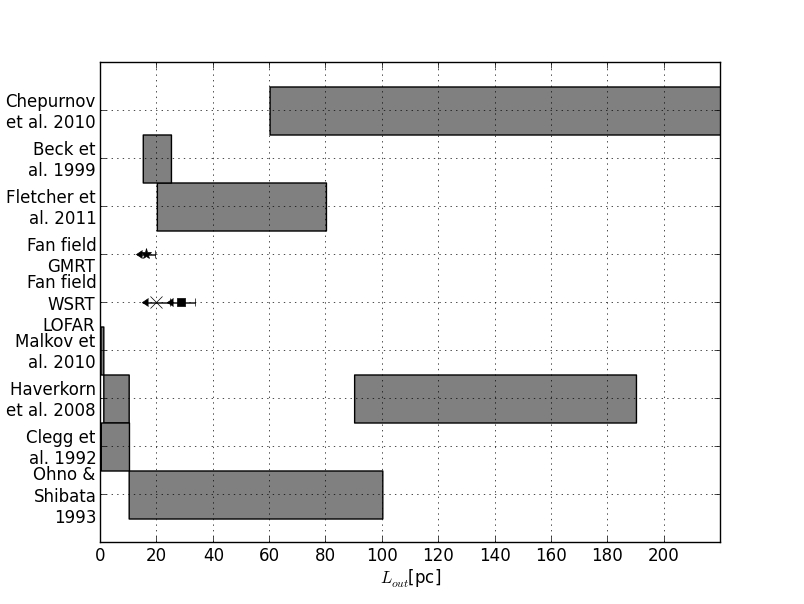}}
\caption{\label{f:outer_scale} Comparison of the LOFAR, WSRT \citep{Bernardi09}, and GMRT \citep{Ghosh12} estimates of the outer scale of turbulence of the Galaxy towards the second Galactic quadrant with earlier observations from \citet{OhnoShibata93}, \citet{Clegg92}, \citet{Haverkorn08}, \citet{Malkov10}, and \citet{Chepurnov10}. The turbulence scale derived by \citet{Fletcher11} and \citet{Beck99} for the nearby spiral galaxies M~51 and NGC~6946 respectively are also shown.}
\end{figure}
\begin{figure}
\resizebox{9cm}{!}{\includegraphics{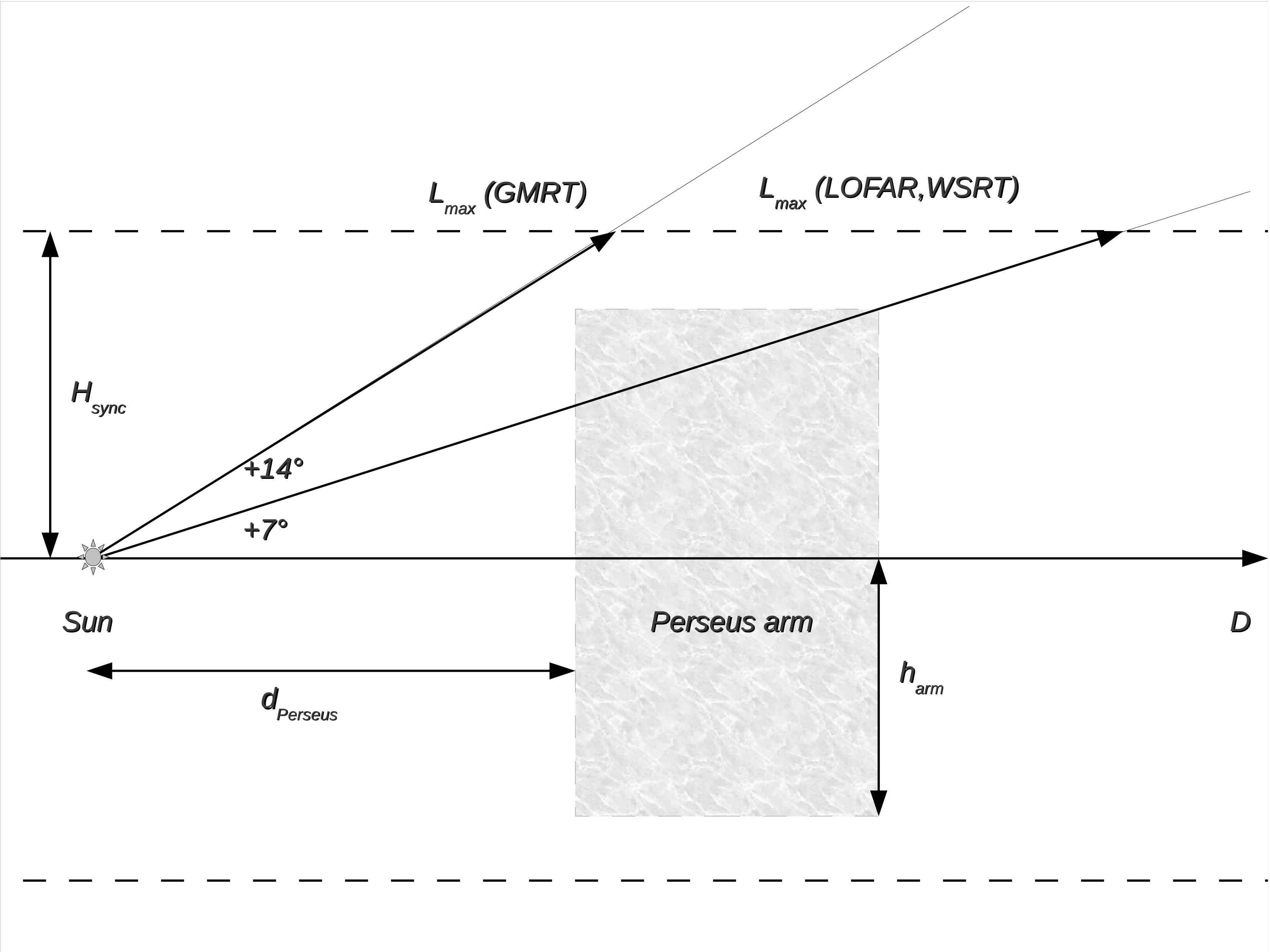}}
\caption{\label{f:model_los} Sketch of the projected spatial configuration for the lines of sight probed by the GMRT, LOFAR, and WSRT observations as a function of distance and Galactic latitude. The Galactic synchrotron scale height ($H_{sync}$) and the Perseus arm scale height ($h_{arm}$) are also shown.}
\end{figure}

These uppers limits are consistent with previous measurements of the outer scale in the Galaxy, as summarised in Fig.~\ref{f:outer_scale} and discussed in the Introduction. In particular, the earlier lower limit for $L_{out}$ of \citet{Wilkinson&Smith74} allows the cell size towards the Fan region to be constrained in the range $\sim 5-20$~pc. While this range of values for $L_{out}$ is inconsistent with the large outer scales found in the Galactic interarms or halo \citep[see e.g.][]{Chepurnov10} and with the small ($L_{out}\lesssim1$~pc) outer scale reported by \citet{Malkov10}, it agrees with the estimate for the spiral arm regions of \citet{Haverkorn08}. This suggests that turbulent fluctuations in the spiral arms dominate the gas density and magnetic field strength along these lines of sight. If so, for a nearest distance to the Perseus arm of $1.95\pm0.04$~kpc \citep{Xu06}, this could indicate that spiral arms would extend at least up to 320~pc from the Galactic disk for the Fan region field or even up to 540~pc for the Ghosh field (Fig.~\ref{f:model_los}). Actually, the complete high-polarisation Fan region stretches out over $(l,b) \approx (90^{\circ}-190^{\circ}, -5^{\circ}-25^{\circ})$ and therefore also encompasses the Ghosh field. Indeed, \citet{Wolleben06} argue from depolarisation arguments that the Fan region has to extend over a range of distances out to the Perseus arm. 
As concerns the disagreement with the small ($L_{out}\lesssim1$~pc) outer scale reported by \citet{Malkov10}, we note that the lower limit of \citet{Wilkinson&Smith74} varies with the assumed total path length. These authors derived the 5~pc lower limits assuming a total path of 500~pc, but lower values are obtained when a deeper origin for the polarised emission \citep{Wolleben06} is considered, thus mitigating the disagreement.

Furthermore, we compare our estimates of the outer scale of magnetic turbulence in the Galaxy to the case of the nearby spiral galaxies M\,51 and NGC\,6946. As noted above, turbulent fluctuations in the spiral arms dominate in our Galaxy and turbulent (and compressed) magnetic fields also play a major role in M\,51, as shown by \citet{Fletcher11} and \citet{Houde13}. However, they find a typical size of 50~pc for the turbulent cells in the magneto-ionic medium of M\,51, about a factor 2 larger than our upper limits towards the outer Galactic spiral arm regions. Such a factor might result from several reasons: a statistical effect because of the averaging over a different sample of probed regions (most of the halo of M\,51 and only three fields for the Galaxy disk), the uncertainties (and assumptions) involved in the estimates and different turbulent parameters in the ISM of the Milky Way and M\,51. The uncertainties involved in the estimates are relevant only for the estimate of \citet{Fletcher11}; thus a difference between the turbulence scale in the Milky Way and in M\,51 may indeed be the major contributor. 
Finally we consider the case of the galaxy NGC\,6946, for which a turbulent scale of about $20\pm10$~pc was estimated by \citet{Beck99}, which is in good agreement with the result of this paper.


We compute the ratio of random to ordered magnetic fields for the WSRT and GMRT data following the procedure above, where the relevant parameters are given in Table~\ref{t:lout_powspec}. The GMRT data claim a tentative $\sim 5\sigma$ detection of diffuse Galactic foreground at a resolution of $10^{\prime}$ and a $\sim 10\sigma$ detection at a resolution of 16$^{\prime}$. We use the latter value of 20~K as a rough approximation of $\sigma_I$ in this field. The numbers quoted for Stokes~I fluctuations are lower limits, since there may be additional fluctuations on scales larger than the ones probed in these studies. However, this will contribute only to more stringent upper limits to the magnetic field ratio.

The resulting ratios of magnetic field components are consistent with earlier estimates in the literature: starlight and synchrotron polarisation data constrain the ratio of regular to turbulent field strengths to $\sim0.6-0.9$ \citep{Beck01,Fosalba01}, and rotation measures of distant pulsars give an even smaller value $\sim0.3$ \citep{OhnoShibata93}. The lower limits indicate that the actual ordered magnetic field component may be even larger, which is not unexpected in the extended Fan region. Due to the extremely high degree of polarization in this region, the regular magnetic field component is believed to dominate over the turbulent component in this field \citep{Haverkorn03b,Wolleben06}. This would account for the deviating ratios of regular to random magnetic field.

In addition, the structure of the Galactic magnetic field affects the motion of CRs across it. The CR electrons are most efficiently scattered and diffused if their gyro radii $r_{g}(E) \simeq 1pc\,(E_{PeV}/Z) \times B_{\mu G}^{-1}$ are similar to the outer scale of magnetic turbulence. For the range of $L_{out}$ values we present in this paper and a total magnetic field strength of $5~\mu$G, CR protons with an energy of 65-130~PeV, which is slightly above the ``knee'', are most efficiently scattered. This is consistent with the idea that the transition from Galactic to extragalactic CRs starts at the ``knee''.

Another consequence of a small outer scale of turbulence in this direction is the possibility of observing anisotropies in gamma-ray flux around young CR sources. \citet{Giacinti12b} describe that young CR sources should emit CRs anisotropically close to the source, up to distances comparable to the outer scale of the turbulent Galactic magnetic field. This should be visible in the gamma-ray flux. For a turbulent outer scale of about 20~pc, this means that these anisotropies would only be visible if the source was located at distances smaller than this, which is very unlikely.




\section{Summary \& Conclusions}
\label{s:conclusion}

In the framework of the commissioning activities to characterise the LOFAR performance, we present results from a LOFAR HBA observation of a field in the Fan region centred at $(l,b) = (137^{\circ}, +7^{\circ})$, between 110 and 174~MHz, using currently available LOFAR data processing software. We show in this paper that fluctuations in the diffuse synchrotron emission can be used to characterise turbulence scales and magnetic field strengths in the ISM. LOFAR is a breakthrough instrument for this work due to its low frequencies (providing high sensitivity for synchrotron emission) and good \textit{uv}-coverage for imaging extended emission. Summarizing the process:

\begin{enumerate}
\item For the first time we detect and image Galactic diffuse synchrotron emission with LOFAR in total intensity, at an angular resolution of about $1\arcmin$ and in a wide frequency range around 160~MHz. \\

\item Data of the target field were carefully calibrated and imaged, and we find faint and complex spatial morphology from the highly polarised Fan region in agreement with the detection by \citet{Bernardi09} using the WSRT. \\

\item Comparing the LOFAR data of the Fan field to the WSRT data, we find the total intensity angular power spectrum of LOFAR in agreement with that of WSRT. Due to the higher resolution of LOFAR, we characterise for the first time the statistical properties of the foreground synchrotron fluctuations as a function of the angular multipole up to $l\sim1300$. The power spectrum of the synchrotron diffuse foreground is approximately a power law up to angular multipoles $\lesssim1300$, corresponding to an angular scale of about $8\arcmin$. The slope we find is in agreement with the one reported by \citet{Bernardi09} within 2 sigma. \\

\item We estimate the outer scale ($L_{out}$) of ISM magnetic turbulence from theoretical arguments that a break in the power spectrum should be observed at a certain critical scale \citep{ChoLazarian02}. We find a range of plausible values in the LOFAR and WSRT data sets and in a third low-frequency field with diffuse synchrotron fluctuations observed with the GMRT of $16-29$~pc. This value is in agreement with previous estimates of outer scales of turbulence in spiral arms, although it is a factor of a few too small to be consistent with outer scale values in the Galactic halo or interarm regions. This suggests that towards the Fan region observed fluctuations are at least in part due to synchrotron emission in the Perseus arm. \\

\item We constrain the ratio for the magnetic field components $B_{o}/B_{r}$ from theoretical estimates of allowed magnetic field strength ratios based on the relative strength of the synchrotron fluctuations with respect to the mean total intensity \citep{Eilek89a,Eilek89b}. Lower limits of the ratio of random to ordered magnetic field strength are found to be 0.3, 0.3, and 0.5 for the three fields considered. These are consistent with magnetic field ratios at other places in the ISM and may indicate a higher than average ordered magnetic field in the Fan region.

\end{enumerate}
Even though the presented LOFAR observations only show a moderate improvement in both resolution and sensitivity over existing WSRT data, they do reveal for the first time the feasibility of imaging interstellar turbulence with LOFAR through fluctuations in synchrotron emission. Furthermore, we prove the usefulness of theoretical estimates of characteristics of interstellar turbulence as applied to these data.

We present indications of a limited accuracy in the flux recovering in the field. This may indicate either that the current LOFAR calibration procedures for complex fields like this one, with extended and point source emission, could not yet be  sufficiently accurate for precise flux calibration or a limited instrumental performance at station level at the time of the observation. We expect that the ongoing technical improvements of the array stations will largely increase sensitivity and decrease artifacts. Moreover, a better accuracy of the recovering of source fluxes may also be achieved by adopting a different observing strategy (e.g.\ using a multi-beam observing mode). 

Future sensitive and high-resolution LOFAR observations in a mosaic mode will allow  a wider portion of the Galaxy to be covered in order to check the spatial dependence of synchrotron fluctuations and the related variation of the turbulent cell size $L_{out}$. Furthermore, mapping the angular power spectrum of synchrotron fluctuations will benefit understanding of the Galactic magnetic field structure, its relation to the turbulence, and the transport of the CR across the magnetised and turbulent ISM. 

\begin{acknowledgements}
The authors thank the anonymous referee for carefully reading the manuscript and providing helpful comments and suggestions in the preparation of the final manuscript. The authors wish to thank M.~Brentjens for his help with the power spectral analysis of LOFAR data analysis. Chiara Ferrari and Giulia Macario acknowledge financial support by the {\it “Agence Nationale de la Recherche”} through grant ANR-09-JCJC-0001-01. LOFAR, the LOw Frequency ARray designed and constructed by ASTRON, has facilities in several countries, which are owned by various parties (each with their own funding sources), and are collectively operated by the International LOFAR Telescope (ILT) foundation under a joint scientific policy. The research leading to these results has received funding from the European Union's Seventh Framework Programme (FP7/2007-2013) under grant agreement number 239490. This work is part of the research programme 639.042.915, which is (partly) financed by the Netherlands Organisation for Scientific Research (NWO). This research has made use of the SIMBAD database, operated at CDS, Strasbourg, France.
\end{acknowledgements}


\begin{thebibliography}{}
\bibitem[Baccigalupi et al.(2001)]{Baccigalupi01} Baccigalupi C., Burigana C., Perrotta F., De Zotti G., La Porta L., Maino D., Maris M., Paladini R. 2001, \aap, 372, 8
\bibitem[Battaner et al.(2009)]{Battaner09} Battaner E., Castellano J. and Masip M. 2009, \apj, 703, 90
\bibitem[Beck et al.(1999)]{Beck99} Beck R., Berkhuijsen E.M. and Uyaniker, B. 1999, in Plasma Turbulence and Energetic Particles in Astrophysics, eds. M. Otrowski and R. Schlickeiser (Krak\'{o}w: Obs. Astron. Univ. Jagiellongski), 5
\bibitem[Beck(2001)]{Beck01} Beck R. 2001, SSR 99, 243
\bibitem[Bernardi et al.(2009)]{Bernardi09} Bernardi G., de Bruyn A.G., Brentjens M.A., et al. 2009, \aap, 500, 965
\bibitem[Bernardi et al.(2010)]{Bernardi10} Bernardi G., de Bruyn A.G., Harker G., et al. 2010, \aap, 522, 67
\bibitem[Beuermann, Kanbach \& Berkhuijsen(1985)]{Beuermann85} Beuermann K., Kanbach G. and Berkhuijsen E.M. 1985, \aap, 153, 17
\bibitem[Bingham \& Shakeshaft(1967)]{Bingham67} Bingham, R.G., \& Shakeshaft, J.R.\ 1967, \mnras, 136, 347 
\bibitem[Brown(2011)]{Brown11} Brown S. 2011, JApA 32, 577
\bibitem[Burkhart et al.(2012)]{Burkhart12} Burkhart B., Lazarian A., Gaensler B.M. 2012, \apj, 749, 145
\bibitem[Carretti et al.(2009)]{Carretti09} Carretti E., Haverkorn M., McConnell D., Bernardi G., Cortiglioni S., McClure-Griffiths N.M., Poppi S. 2009, RevMexAA (SC) 36, 9
\bibitem[Chepurnov et al.(2010)]{Chepurnov10} Chepurnov A., Lazarian A., Staminirovi\'{c} S., Heiles C., \& Peek, J.E.G. 2010, \apj, 714, 1398
\bibitem[Chepurnov \& Lazarian(2010)]{ChepurnovLazarian10} Chepurnov A., Lazarian A. 2010, \apj, 710, 853
\bibitem[Cho \& Lazarian(2002)]{ChoLazarian02} Cho J., Lazarian A. 2002, \apj, 575, 63
\bibitem[Cho \& Lazarian(2003)]{ChoLazarian03} Cho J. \& Lazarian A. 2003, \mnras, 345, 325
\bibitem[Cho \& Lazarian(2010)]{ChoLazarian10} Cho J., Lazarian A. 2010, \apj, 720, 1181
\bibitem[Clegg et al.(1992)]{Clegg92} Clegg, A.W., Cordes, J.M., Simonetti, J.M., \& Kulkarni, S.R.\ 1992, \apj, 386, 143 
\bibitem[Cornwell et al.(2005)]{Cornwell05} Cornwell T.J., Golap K. \& Bhatnagar S. 2005, ASP Conf. Ser. 347, 86
\bibitem[Cornwell et al.(2008)]{Cornwell08} Cornwell T.J., Golap K. \& Bhatnagar S. 2008, IEEE J. Selected Topics in Signal Processing 2, 647
\bibitem[de Oliveira-Costa et al.(2008)]{de Oliveira-Costa08} de Oliveira-Costa A., Tegmark M., Gaensler B. M., Jonas J., Landecker T. L., Reich P. 2008, \mnras, 388, 247
\bibitem[Eilek(1989a)]{Eilek89a} Eilek J.A., AJ 98, 244 (Paper a)
\bibitem[Eilek(1989b)]{Eilek89b} Eilek J.A., AJ 98, 256 (Paper b)
\bibitem[Elmegreen \& Scalo(2004)]{Elmegreen04} Elmegreen B.G. \& Scalo J. 2004, \araa, 42, 211
\bibitem[Fosalba et al.(2001)]{Fosalba01} Fosalba P., Lazarian A., Prunet S. \& Tauber J.A. 2001, \apj, 564, 762
\bibitem[Fletcher et al.(2011)]{Fletcher11} Fletcher A., Beck R., Shukurov A., Berkhuijsen E.M., Horellou C.\ 2011, \mnras, 412, 2396
\bibitem[Gaensler et al.(2011)]{Gaensler11} Gaensler B.M., Haverkorn M., Burkhart B., et al.\ 2011, \nat, 478, 214 
\bibitem[Giacinti \& Sigl(2012a)]{Giacinti12a} Giacinti G. \& Sigl G. 2012, PhRvL 109, 1101 (Paper a)
\bibitem[Giacinti et al.(2012b)]{Giacinti12b} Giacinti G., Kachelrie\ss\,M., Semikoz D.V. 2012, PhRvL 108, 1101 (Paper b)
\bibitem[Giardino et al.(2001)]{Giardino01} Giardino G., Banday A.J., Fosalba P., G\'{o}rski K.M., Jonas J.L., O'Mullane W., Tauber J. 2001, A\&A 371, 708
\bibitem[Ghosh et al.(2012)]{Ghosh12} Ghosh A., Prasad J., Bharadwaj S., Ali S.S., Chengalur J.N. 2012, \mnras, 426, 3295
\bibitem[Goldreich \& Sridhar(1995)]{Goldreich&Sridhar95} Goldreich P. \& Sridhar S. 1995, \apj, 438, 763
\bibitem[Haslam et al.(1982)]{Haslam82} Haslam C.G.T., Salter C.J., Stoffel H. \& Wilson W.E. 1982, \aap, S 47, 1
\bibitem[Haverkorn et al.(2008)]{Haverkorn08} Haverkorn M., Brown J.C., Gaensler B.M. and McClure-Griffiths N.M. 2008, \apj 680, 362
\bibitem[Haverkorn et al.(2003a)]{Haverkorn03a} Haverkorn, M., Katgert, P., \& de Bruyn, A.G.\ 2003, \aap, 403, 1045, Paper a
\bibitem[Haverkorn et al.(2003b)]{Haverkorn03b} Haverkorn, M., Katgert, P., \& de Bruyn, A.G.\ 2003, \aap, 404, 233, Paper b
\bibitem[Hamaker et al.(1996)]{Hamaker96} Hamaker J.P., Bregman J.D. \& Sault R.J. 1996, \aaps, 117, 137
\bibitem[Heald et al.(2010)]{Heald10} Heald G., McKean J., Pizzo R., van Diepen G., van Zwieten J.E., et al. 2010, PoS(ISKAF2010), 57
\bibitem[Heald et al.(2011)]{Heald11} Heald G., Bell M.R., Horneffer A. et al. 2011, JApA 32, 589
\bibitem[Hill et al.(2008)]{Hill08} Hill, A.S., Benjamin, R.A., Kowal, G., et al. 2008, \apj, 686, 363 
\bibitem[Houde et al.(2013)]{Houde13} Houde M., Fletcher A., Beck R., Hildebrand R.H., Vaillancourt J.E., Stil J.M. 2013, \apj, 766, 49
\bibitem[Iacobelli et al.(2013)]{Iacobelli13} Iacobelli M., Haverkorn M. and Katgert P. 2013, \aap, 549, 56
\bibitem[Junklewitz et al.(2011)]{Junklewitz11} Junklewitz H. and En{\ss}lin T.~A. 2011, \aa, 530, 88
\bibitem[Kogut(2012)]{Kogut12} Kogut A. 2012, ApJ 753, 110
\bibitem[Landecker \& Wielebinski(1970)]{LandeckerWielebinski70} Landecker T.L. \& Wielebinski R. 1970, AuJPA, 16, 1
\bibitem[Lazarian \& Pogosyan(2012)]{Lazarian12} Lazarian A. and Pogosyan D. 2012, \apj, 747, 5
\bibitem[Lazarian \& Shutenkov(1990)]{LazarianShutenkov90} Lazarian A. \& Shutenkov V.P. 1990, PAZh, 16, 690 (translated Sov. Astron. Lett., 16, 297)
\bibitem[La Porta et al.(2008)]{LaPorta08} La Porta L., Burigana C., Reich W., Reich P. 2008, \aap, 479, 641
\bibitem[Kazemi et al.(2011)]{Kazemi11} Kazemi S., Yatawatta S., Zaroubi S., et al. 2011, \mnras, 414, 1656
\bibitem[Mac Low(2004)]{MacLow04} Mac Low M.-M. 2004, Astrophys. Space Sci. 289, 323 
\bibitem[Malkov et al.(2010)]{Malkov10} Malkov M.A., Diamond P.H., Drury L., and Sagdeev R. Z. 2010, ApJ 721, 750
\bibitem[Nota \& Katgert(2010)]{NotaKatgert10} Nota T. \& Katgert P. 2010, \aap, 513, 65
\bibitem[Offringa et al.(2010)]{Offringa10} Offringa A.R., de Bruyn A.G., Biehl M., Zaroubi S., Bernardi G. \& Pandey V.N. 2010, \mnras, 405, 155
\bibitem[Offringa et al.(2012)]{Offringa12} Offringa A.R., van de Gronde J.J. \& Roerdink J.B.T.M. 2012, \aap, 539, A95
\bibitem[Ohno \& Shibata(1993)]{OhnoShibata93} Ohno H. \& Shibata S. 1993, \mnras, 262, 953
\bibitem[Page et al.(2007)]{Page07} Page L., Hinshaw G., Komatsu E., et al. 2007, ApJS 170, 335
\bibitem[Pandey et al.(2009)]{Pandey09} Pandey V.N., van Zwieten J.E., de Bruyn A.G., Nijboer R. 2009, ASPC 407, 384
\bibitem[Pizzo et al.(2010)]{Pizzo10} Pizzo R.F. et al. 2010, ``The Lofar Imaging Cookbook'', internal ASTRON report 
\bibitem[Regis(2011)]{Regis11} Regis M. 2011, APh 35, 170
\bibitem[Scalo \& Elmegreen(2004)]{Scalo04} Scalo J. \& Elmegreen B.G. 2004, \araa, 42, 275
\bibitem[Seljak(1997)]{Seljak97} Seljak U. 1997, \apj, 482, 6
\bibitem[Sotomayor et al.(2013)]{Sotomayor13} Sotomayor-Beltran C.,  Sobey C., Hessels J.W.T., de Bruyn G. et al. 2013, \aap, 552, 58
\bibitem[Stepanov et al.(2012)]{Stepanov12} Stepanov R. Shukurov A., Fletcher A., Beck R., La Porta L., Tabatabaei F.S. 2012, arXiv 1205.0578S
\bibitem[Tasse et al.(2013)]{Tasse13} Tasse C., van der Tol S., van Zwieten J., van Diepen G., Bhatnagar S. 2013, \aap, 553, 105
\bibitem[Tegmark(1997)]{Tegmark97} Tegmark M. 1997, PRD 56, 8
\bibitem[Turtle et al.(1962)]{Turtle62} Turtle A.J., Pugh J.F., Kenderdine S., Pauliny-Toth I.I.K. 1962, \mnras, 124, 297
\bibitem[Verschuur(1968)]{Verschuur68} Verschuur, G.~L.\ 1968, The Observatory, 88, 15 
\bibitem[Waelkens et al.(2009)]{Waelkens09} Waelkens A.~H., Schekochihin A.~A. and En{\ss}lin T.~A. 2009, \mnras, 398, 1970
\bibitem[Wieringa et al.(1993)]{Wieringa93} Wieringa M.H., de Bruyn A.G., Jansen D., Brouw W.N. and Katgert P. 1993, \aap, 268, 215
\bibitem[Wilkinson \& Smith(1974)]{Wilkinson&Smith74} Wilkinson A., Smith F.G. 1974 \mnras, 167, 593
\bibitem[Wolleben et al.(2006)]{Wolleben06} Wolleben, M., Landecker, T.~L., Reich, W., \& Wielebinski, R.\ 2006, \aap, 448, 411 
\bibitem[Xu et al.(2006)]{Xu06} Xu, Y., Reid, M.J., Zheng, X.W., \& Menten, K.M.\ 2006, Science, 311, 54 
\\
\end{thebibliography}
\end{document}